\documentclass[prx,twocolumn,floatfix,10pt,superscriptaddress,longbibliography]{revtex4-1}
\usepackage[utf8]{inputenc}

\usepackage{natbib}
\usepackage{float}
\usepackage{placeins}
\usepackage{graphicx}
\usepackage{latexsym}
\usepackage{amssymb}
\usepackage{amsmath}
\usepackage{amsfonts}
\usepackage{gensymb}
\usepackage{bm}
\usepackage{siunitx}
\usepackage{hyperref}
\usepackage{braket}
\usepackage{bbold}
\usepackage[caption=false]{subfig}
\usepackage[dvipsnames]{xcolor}
\usepackage[normalem]{ulem}
\usepackage[version=4]{mhchem}

\usepackage{titletoc}

\definecolor{pal0}{rgb}{0.8941, 0.102 , 0.1098}
\definecolor{pal1}{rgb}{0.2157, 0.4941, 0.7216}
\definecolor{pal2}{rgb}{0.302 , 0.6863, 0.2902}
\definecolor{pal3}{rgb}{0.5961, 0.3059, 0.6392}
\definecolor{pal4}{rgb}{1.    , 0.498 , 0.    }

\newcommand{\E}{\mathbb E}

\hypersetup{
  pdfauthor = {},
  pdftitle = {}
}

\begin{document}

\title{Information-theoretic entropy imaging of quantum matter}

\author{Chen Wu}
\affiliation{Department of Physics, University of California, San Diego, CA 92093, USA}

\author{Leonard W. Cao}
\affiliation{Department of Physics, University of California, San Diego, CA 92093, USA}

\author{Lingyuan Lyu}
\affiliation{Department of Physics, University of California, San Diego, CA 92093, USA}

\author{Liam Cohen}
\affiliation{Department of Physics, University of California, Santa Barbara, CA 93106, USA}

\author{Jiahui Nie}
\affiliation{Department of Physics, University of California, Berkeley, CA 94720, USA}
\affiliation{Materials Sciences Division, Lawrence Berkeley National Laboratory, Berkeley, CA 94720, USA.}
\affiliation{Graduate Group in Applied Science and Technology, University of California, Berkeley, CA 94720, USA.}

\author{Can Uzundal}
\affiliation{Department of Physics, University of California, Berkeley, CA 94720, USA}
\affiliation{Materials Sciences Division, Lawrence Berkeley National Laboratory, Berkeley, CA 94720, USA.}
\affiliation{Department of Chemistry, University of California at Berkeley, Berkeley, California 94720, USA.}

\author{Yujia Hu}
\affiliation{Department of Physics, University of California, Berkeley, CA 94720, USA}

\author{Quanjin Wang}
\affiliation{Department of Physics, University of California, San Diego, CA 92093, USA}

\author{Ziying Yan}
\affiliation{Department of Physics, Princeton University, Princeton, NJ 08544, USA}

\author{ Sneh Pancholi}
\affiliation{Department of Physics, University of Oxford, Oxford, OX1 2JD, UK}

\author{Kenji Watanabe}
\affiliation{Research Center for Electronic and Optical Materials, National Institute for Materials Science, Namiki 1-1, Tsukuba, Ibaraki 305-0044, Japan}

\author{Takashi Taniguchi}
\affiliation{Research Center for Materials Nanoarchitectonics, National Institute for Materials Science, Namiki 1-1, Tsukuba, Ibaraki 305-0044, Japan}

\author{Feng Wang}
\affiliation{Department of Physics, University of California, Berkeley, CA 94720, USA}
\affiliation{Materials Sciences Division, Lawrence Berkeley National Laboratory, Berkeley, CA 94720, USA.}
\affiliation{Kavli Energy NanoScience Institute, University of California Berkeley and Lawrence Berkeley National Laboratory, Berkeley, CA 94720, USA.}

\author{ Andrea F. Young}
\affiliation{Department of Physics, University of California, Santa Barbara, CA 93106, USA}

\author{Tarun Grover}
\affiliation{Department of Physics, University of California, San Diego, CA 92093, USA}

\author{ Monica T. Allen}
\affiliation{Department of Physics, University of California, San Diego, CA 92093, USA}

\date{October 24, 2026}

\begin{abstract}

    Phase transitions can generate intricate spatial organization invisible to spatially averaged probes, yet quantitative measures of this organization remain limited. Here we develop ``entropy imaging,'' an information-theoretic technique that harnesses lossless compression to quantify spatial complexity, predictability, and order in quantum materials. Using this approach, emergent spatial organization can be imaged and converted into an experimentally-accessible observable: the computable information density (CID), a compression-based measure of     information complexity that approaches the Shannon entropy rate in the large-system limit. Applied to millikelvin microwave impedance microscopy of graphene, the method reveals a pronounced maximum in the CID near quantum Hall transitions, a quantitative measure of the information-rich spatial organization of the critical landscape. A disordered Hofstadter model reproduces this behavior and shows that the CID can identify the quantum critical point from a sequence of images, without ensemble averaging over disorder realizations. The same framework tracks fractional quantum Hall transitions and resolves isolated fractional puddles that coexist with a disordered background. By contrast, the trivial band-edge crossover in a MoSe$_2$/WS$_2$ moiré heterobilayer exhibits qualitatively different behavior, suggesting that the pronounced CID enhancement observed at the quantum Hall critical point is not simply a generic feature of a conductive-insulating crossover. We further show that the geometric information encoded in CID is not reducible to a correlation length, and it can even distinguish spatial textures with identical multifractal statistics. Together, these results establish lossless compression as a framework for quantifying emergent spatial organization in quantum materials, with portability across imaging modalities and physical systems.
\end{abstract}

\maketitle

Microscopy has made states of matter visible as spatial objects, revealing where domains nucleate, how phase boundaries move, and when conducting pathways connect. Near quantum Hall transitions, for example, disorder produces a network of conductive filaments whose shapes and connections evolve nontrivially across the quantum critical point \cite{goswami2007quantum,pereira2011correlated,Halperin2013,ding2022correlated,hashimoto2008stm,martin2008observation,Cui2016Gr,uri2020sot,kim2021afm,wang2021imaging,zhou2025MIM}. Spatial averages and global probes can identify phase transitions but not the geometric arrangements through which they unfold \cite{martin2008observation}.

This raises a fundamental question:
how can the spatial complexity of a physical system be quantified from a single image?
In physics, entropy provides a standard measure of disorder and information; however, it is defined statistically over an ensemble of configurations and does not by itself specify the geometric organization of an individual snapshot \cite{shannon1948mathematical}. Existing image-based metrics capture selected aspects of that organization: correlation lengths report a characteristic scale but not the full morphology, while cluster statistics quantify connectivity but require a binarization threshold \cite{Chalker1988,cilibrasi2005clustering,Deshpande2009}.
Multifractal analysis, central for studying localization physics, measures how weight is distributed across scales but not the full geometric arrangement \cite{mandelbrot1979fractals,halsey1986fractal,huckestein1995scaling,obuse2008boundary,evers2008multifractality}, precisely the information that distinguishes a connected percolating network from a set of isolated puddles \cite{Chalker1988}.

We reframe this problem through the lens of algorithmic information theory, which quantifies the complexity of
an \textit{individual} object without reference to an ensemble of possible outcomes, departing from the probabilistic concept of entropy.
In the 1960s, mathematicians Solomonoff, Kolmogorov, and Chaitin developed the idea that the complexity of a sequence, such as an image encoded as a string of symbols, can be defined as the length of the shortest description or program that reproduces it exactly~\cite{solomonoff1964formal, kolmogorov1965three, chaitin1966length}. Ordered or repetitive patterns admit short descriptions, while intricate, unpredictable ones do not.
Although this Kolmogorov complexity is typically uncomputable, lossless data compression algorithms turn this concept into an experimentally accessible quantity: the computable information density (CID), defined as the ratio of an image's compressed to original length~\cite{ziv1977universal,ziv1978compression}.
Remarkably, for statistically translation-invariant and self-averaging sources, this quantity approaches the Shannon entropy rate in the large system limit,
establishing a bridge between algorithmic complexity and the probabilistic notion of entropy~\cite{wyner1991fixed, wyner1994sliding, ornstein2002entropy}.
\textcolor{black}{Building on this connection, Martiniani, Chaikin, and Levine showed that the CID can identify hidden order and  phase transitions in systems of active Brownian particles, even without prior knowledge of the relevant order parameter ~\cite{martiniani2019quantifying}. More recently, it was shown that CID can also detect quantum phase transitions~\cite{chen2025zipping}.}
\textcolor{black}{Throughout this paper, we use `entropy' in this information-theoretic sense rather than to denote thermodynamic entropy. When applied to experimental data, CID should be viewed as an operational compression-based measure of information complexity, with the connection to entropy rate becoming exact only under the appropriate asymptotic conditions.}

This connection is particularly well-matched to microscopy, which supplies single snapshots of a particular configuration, not a full probability distribution. CID analysis therefore begins from the object the experiment actually provides: a real-space image.
Because compression-based measures of algorithmic complexity have not previously been applied to real-space imaging of quantum materials, we deliberately benchmark the approach in systems whose underlying physics is independently established.
To put this framework to an experimental test, we use millikelvin microwave impedance microscopy (MIM) to track evolving electronic landscapes across integer and fractional quantum Hall transitions in graphene \cite{barber2022microwave,cao2023millikelvin,yan2026microwave,dean2011fqhe}
and a trivial band-edge crossover in a MoSe$_2$/WS$_2$ moiré heterobilayer \cite{Carr2020NatRevMatMethods,Andrei2021NatRevMatMarvels,Nuckolls2024AMP,zhai2025twistronics,lu2025bosonic,li2021mose2ws2}. In graphene, the CID develops a pronounced maximum near the quantum Hall critical point as extended conductive filaments emerge, a behavior reproduced by a disordered Hofstadter model.
\textcolor{black}{The CID likewise tracks fractional quantum Hall transitions, and a spatially resolved implementation detects isolated fractional puddles that coexist with a disordered background. By contrast, a trivial band-edge crossover lacks the CID enhancement observed at the quantum critical point, indicating that the effect is not a generic consequence of entering a gapped state.}
We further show that CID captures spatial information not reducible to conventional multifractal statistics or correlation length alone.

\textcolor{black}{To test the portability of this framework, we apply the same compression-based approach to data acquired with other imaging modalities, including ultrafast electron microscopy, scanning transmission X-ray microscopy, atomic force microscopy, and optical microscopy. Applied retroactively to existing datasets, the CID tracks evolving geometric complexity in patterns ranging from field-driven magnetic skyrmion textures to wrinkle networks in growing bacterial biofilms. Together, these results establish entropy imaging as an experimentally accessible probe of emergent spatial organizations in complex systems, from quantum materials to living matter.}

\begin{figure*}[h!btp]
    \includegraphics[width=1.0\textwidth]{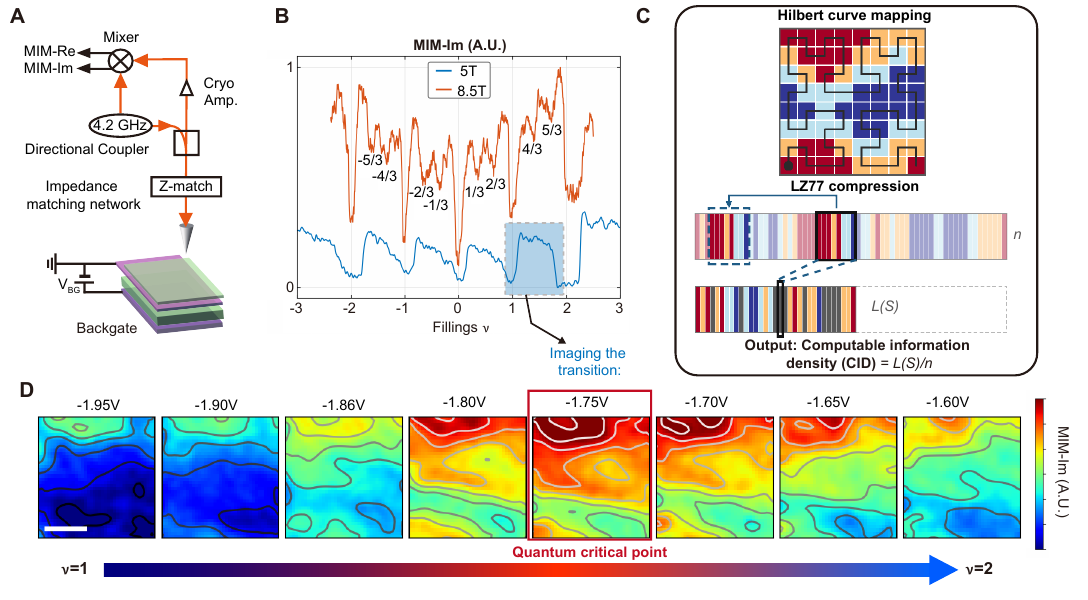}
    \caption{\label{fig:Schematics}
    \textbf{Real-space imaging and compression analysis of a quantum Hall transition.
    }
    \textbf{(A)} Schematic of the microwave impedance microscopy (MIM) measurement setup, in which a GHz excitation is coupled to a metallic tip to probe the local electronic compressibility of gated monolayer graphene.
    \textbf{(B)} Fixed-position MIM-Im gate sweeps between $\nu = -2 $ and 2 at $B = \SI{5}{T} $ (blue) and $B = \SI{8.5}{T} $ (red). The shaded region marks the $\nu = 1 \to 2$ transition imaged in panel (D).
    \textbf{(C)}
    Compression analysis workflow: Each MIM$(x,y)$ image (top panel) is serialized along a Hilbert space-filling curve (black line) and compressed using the LZ77 algorithm, in which repeated patterns are encoded as back-references to earlier parts of the sequence (blue arrow in the middle panel).
    The computable information density (CID) is defined as $CID = L(S)/n$, where $L(S)$ is the compressed bit length and $n$ is the length of the original sequence (bottom panel).
    The compression is lossless and based on the length-matching algorithm.
    \textbf{(D)}
    Representative MIM-Im images across the $\nu = 1 \to 2$ transition at $\SI{5}{T}$, showing the evolution of the electronic landscape near the critical point at $\SI{-1.75}{V}$. The scale bar is $1\mu m$.
    }

\end{figure*}

 \begin{figure*}[h!tbp]
     \includegraphics[width=0.8\textwidth]{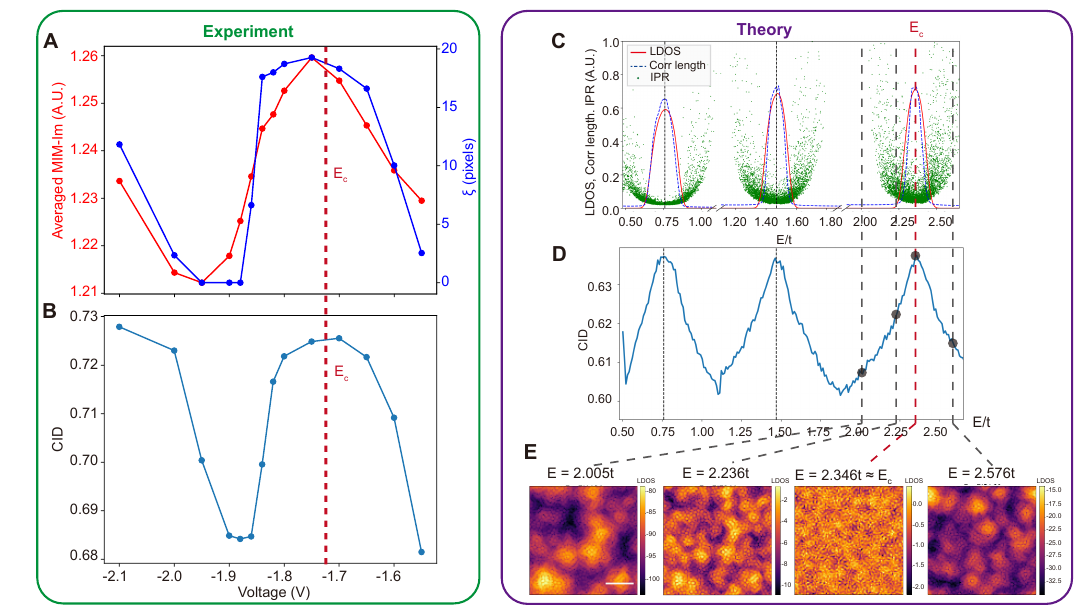}
     \caption{\label{fig:CID}
     \textbf{Spatial information complexity peaks at the quantum Hall transition.}
     \textbf{(A-B)} Experimental analysis of the $\nu=1 \to 2$ transition.
     \text{(A):} Spatially averaged MIM-Im signal (red, left axis) and the correlation length $\xi$ extracted from
     the same image series (blue, right axis) versus gate voltage. Both are strongly enhanced near the transition.
     \text{(B):} CID calculated from the identical image series. The CID develops a pronounced maximum near $E_{\mathrm{c}}$, coincident with the critical point $E_c$.
     \textbf{(C-E)} Disordered Hofstadter model calculation.
     \text{(C):} Spatially averaged local density of states (LDOS, red), correlation length (blue dashed), and inverse participation ratio (IPR, green) as functions of energy.
    \text{(D):} CID calculated from the corresponding LDOS maps, exhibiting maxima at the transitions.
    \text{(E):} Representative LDOS maps at the energies indicated by the black dashed lines in (C,D), illustrating the evolution of the spatial texture near
    $E = 2.346t \approx E_{\mathrm{c}}$ (red dashed line). The scale bar is approximately 20$\l_B$.
     }

 \end{figure*}

\begin{center}
\textbf{RESULTS}
\end{center}

\textbf{I. Mapping spatial complexity across a quantum Hall transition in graphene.}
\textcolor{black}{
To benchmark entropy imaging,
we use millikelvin microwave impedance microscopy (mK MIM) to track the real-space evolution of the quantum critical landscape near Landau level transitions in graphene, as illustrated schematically in Fig.~\ref{fig:Schematics}A.}
Throughout this work, we analyze the imaginary component of the tip–sample admittance, Im[$Y_\textrm{ts}$], which is related to the local electronic compressibility $Y_\textrm{ts} \propto -i \frac{dn}{d \mu} $ to leading order, where $n$ is the local carrier density and $\mu$ is the chemical potential \cite{Taige2023comsol}. A microwave signal at \SI{4.2}{GHz} is delivered via coaxial cables to the apex of the metallic scanning probe, and the reflected microwave signal is amplified and demodulated to extract the complex tip-sample admittance.
As the carrier density in the system is tuned using a gate voltage $V_g$, successive Landau level transitions produce alternating regions of high and low MIM response (Fig.~\ref{fig:Schematics}B) \cite{Cui2016Gr,Cao2025layer}.

\begin{figure*}[h!tbp]
    \includegraphics[width=0.9\textwidth]{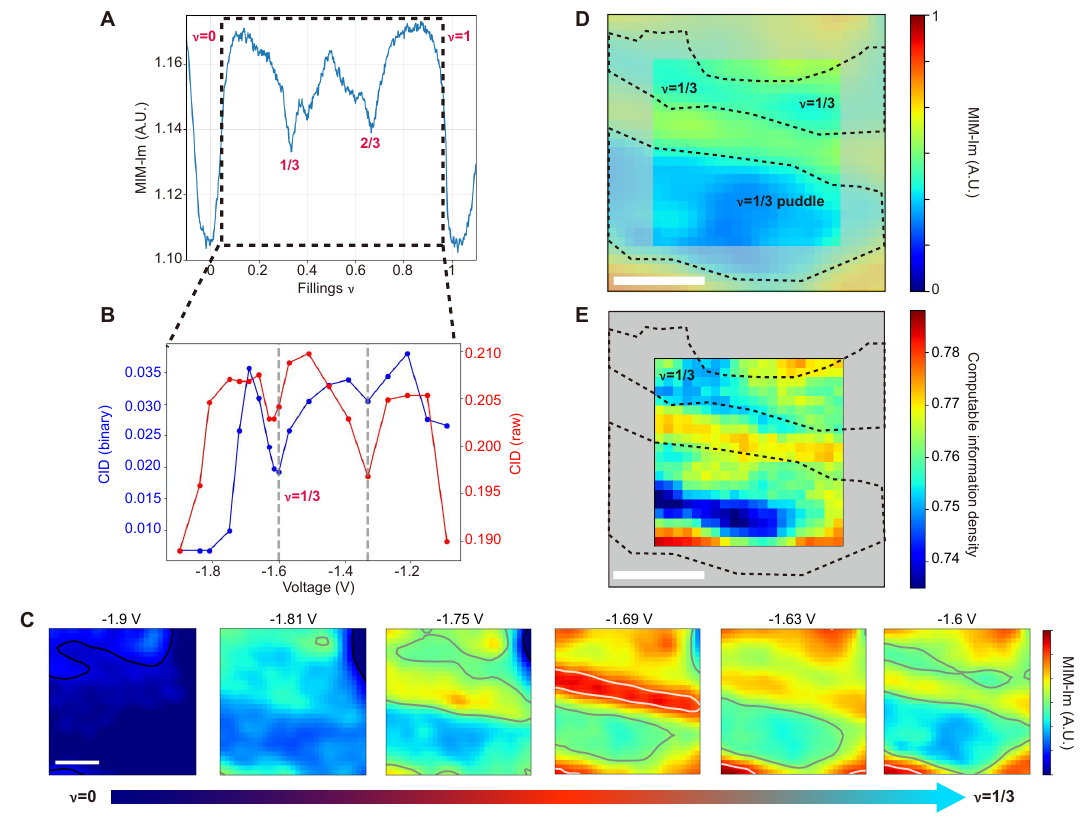}
    \caption{\label{fig:Simulation}
    \textbf{Compression-based entropy imaging of fractional quantum Hall states in graphene.}
    \textbf{(A)} Fixed-position MIM-Im gate sweep at $B=8.5$ T spanning $0<\nu<1$, with incompressible features at $\nu=1/3$ and $2/3$.
    \textbf{(B)} CID extracted from a series of MIM images using the continuous MIM signal (red) and a threshold-defined binary representation (blue). The CID is enhanced in the transition regions and suppressed upon entering the incompressible fractional quantum Hall (FQH) states.
    \textbf{(C)} Series of MIM-Im(x,y) images spanning the $\nu = 0 \to 1/3$, showing the real-space evolution of the conductivity landscape. The scale bar is equivalent to $\SI{1}{\mu m}$.
    \textbf{(D)} MIM-Im image near $\nu = 1/3$.
    \textbf{(E)} The spatial CID(x,y) calculated by convolving a $10\times10$ box across the image in (D). The map resolves variations in spatial complexity and local formation of FQH puddles (blue regions).
    }

\end{figure*}

To investigate the transition in real space, we acquired a series of real-space MIM$(x,y)$ images spanning the  $\nu = 1 \to 2 $ Landau level transition (Fig.~\ref{fig:Schematics}D). Each image covers a $3 \times 3$  $\mu m^2$ area with a \SI{75}{nm} pixel size, comparable to the spatial resolution of the probe.
At 5 T, this field of view spans approximately 260 magnetic lengths.
This large scale separation is particularly important near the quantum Hall critical point, where the localization length diverges and a limited imaging window can truncate the extended critical landscape.
Similar evolution was observed in a second region of the sample and near fractional filling factors (Fig.~\ref{fig:Simulation}C), demonstrating that the spatial patterns are reproducible and are not specific to a particular scan location.

\textit{\textbf{\textcolor{black}{From a single image to spatial information complexity.}}}
Consider a microscopy image,
consisting of an $M \times N$ grid of pixels.
The Shannon information entropy, $H_\textrm{Shan}(X) = - \sum_{x_i} P(x_i) \log P(x_i)$,
is defined over the probability distribution of all possible  configurations, $x_i$, of an $M \times N$ grid of pixels \cite{shannon1948mathematical, cover1991information}.
Reconstructing this distribution experimentally would require repeating measurements over an astronomically large ensemble, whereas microscopy ordinarily provides only a single configurational snapshot.

Kolmogorov complexity instead quantifies the information content of an \textit{individual} configuration.
Upon flattening the image into a string $S$ of length $n = MN$, we define the Kolmogorov complexity $K(S)$ as the length of the shortest program that outputs $S$ exactly \cite{li1989kolmogorov,cover1991information,li2008introduction}.
$K(S)$ therefore quantifies spatial complexity: patterns containing regularity or repeated motifs admit shorter descriptions, while disordered images correspond to near-maximal complexity. Although Kolmogorov complexity is generally uncomputable, it is closely connected to the object central to our discussion, namely, the CID. Consider representing a single image as a symbolic sequence $X_1,X_2,\ldots,X_n\equiv X_1^n$, and suppose that in the large-system limit it is a typical realization of a stationary ergodic process. The Shannon entropy rate,
$h=\lim_{n\to\infty}\frac{1}{n}H(X_1^n)$,
quantifies the irreducible information per symbol after correlations are taken into account~\cite{cover1991information}. The Kolmogorov complexity density satisfies $K(X_1^n)/n\to h$ in the large-$n$ limit for almost every realization~\cite{brudno1983entropy}, while universal lossless compression schemes such as Lempel--Ziv achieve the same asymptotic rate, the compressed length $L(X_1^n)$ obeys $L(X_1^n)/n\to h$ ~\cite{Wyner1989asymptotic,ziv1977universal,ziv1978compression}. Thus, under stationarity and ergodicity, CID provides a computable estimate of both the algorithmic-complexity density and the Shannon entropy rate, i.e., the optimal lossless compression rate, from a single, large enough image.

\textit{\textbf{Implementation.}}
Each MIM$(x, y)$ image is mapped to a one-dimensional sequence via a space-filling Hilbert curve, which preserves spatial locality, and losslessly compressed with LZ77, wherein repeated motifs are replaced by back-references to earlier occurrences (blue arrow in the middle panel of Fig.~\ref{fig:Schematics}C); the computable information density is the ratio $CID = L(S)/n$ of the compressed bit length $L(S)$ to the length n of the original sequence \cite{hilbert1935stetige,moon2001analysis,ziv1977universal,martiniani2019quantifying}.
Low values denote a highly ordered, easily compressed pattern; high values denote a complex, information-rich one.
\textcolor{black}{For finite experimental images, however, we treat CID as a compression-based estimator of spatial information complexity rather than as a measurement of thermodynamic entropy.}

Fig.~\ref{fig:CID}B shows the measured CID across the $\nu = 1 \to 2$  transition, extracted from a sequence of MIM images collected as a function of gate voltage (shown in Fig.~\ref{fig:Schematics}D).  All encoding choices, including normalization, alphabet, serialization, and compressor are held fixed across the experimental series.
Because the absolute CID value depends on these choices and on image size, we interpret relative changes within a fixed series rather than comparing absolute values across unrelated datasets.
\textcolor{black}{
One source of this size dependence is finite-sequence compression overhead, which can shift the absolute CID. Because all images within a given series contain the same number of pixels and are processed identically, this contribution is approximately constant across the series; our analysis therefore focuses on the evolution of CID with the tuning parameter rather than its absolute magnitude. Coarse-graining tests confirm that the characteristic transition-associated trends remain robust over the relevant range of image sizes and resolutions (see Supplementary Information).}
\textcolor{black}{Importantly, the CID serves as a model-agnostic diagnostic of changes in spatial organization rather than as a standalone identifier of a particular phase or critical point, allowing broader application to different physical systems and materials.
In an unfamiliar system, a pronounced maximum, minimum, or abrupt change in CID across a tuning parameter would flag a regime of unusual spatial reorganization, motivating comparison with conventional observables and microscopic models.}

We find that the CID is low near integer fillings, where the MIM images show uniform insulating regions interrupted by sparse, isolated conductive islands.
\textcolor{black}{
Approaching the independently identified ($\nu=1 \rightarrow 2$) transition, CID develops a pronounced local maximum that coincides with the emergence of the filamentary texture (Fig.~\ref{fig:Schematics}D).
Because CID responds to the full spatial texture, additional variations can occur outside the immediate critical window; throughout this work, we associate a CID feature with quantum Hall criticality only when it coincides with a Landau level transition identified independently from conventional observables, such as the spatially averaged MIM response.}
The correlation length $\xi$, extracted from binarized MIM images using a second-moment estimator, is also strongly enhanced near the transition (Fig.~\ref{fig:CID}A).
For images of finite size,
$\xi$ is bounded by the finite field of view, so we interpret this as a strong enhancement of spatial correlations accompanying the transition rather than as a direct observation of a critical divergence.
The CID, however, is not simply a monotonic repackaging of $\xi$:
the correlation length characterizes the spatial range of correlations, whereas CID is sensitive to broader geometric organization.
Their coincidence near the quantum Hall transition is an empirical property of the critical landscape rather than a definitional consequence of compression.
One counter-example is the site-percolation model (discussed in the Supplementary Information and Fig.~\ref{fig:CID_sitepercolation}), which makes this distinction explicit.

The complexity that emerges at the critical point cannot be explained simply by the balance of high- and low-response pixels alone. To demonstrate this, we reshuffled the pixel positions of each image to eliminate all spatial correlations and then compressed the randomized dataset, yielding ${CID}_\textrm{rand}$ (See SI). The ratio of the measured CID to that of its reshuffled counterpart ($\textrm{normalized } CID={CID}_\textrm{data}/{CID}_\textrm{rand}$) remains sharply peaked at the transition, rising above its off-critical value (Fig.~\ref{fig:Supp_MIM_shuffled}(a)). The same is true at the fractional transitions (Fig. ~\ref{fig:Supp_MIM_shuffled}(b)).
\textcolor{black}{
Because pixel shuffling preserves the histogram of response values while destroying spatial correlations, the robust peak in the ratio ${CID}_\textrm{data}/{CID}_\textrm{rand}$ at the transition isolates a contribution from the spatial organization of the critical landscape. The normalized CID therefore distinguishes changes in the geometry of the landscape from changes in the image histogram alone.}

The observed behavior is consistent with established models, most notably Chalker–Coddington \cite{Chalker1988}, which predict the emergence of an intricate, scale-spanning conducting network precisely at the critical energy. Quantum Hall criticality has been characterized largely using localization scaling and multifractal statistics, rather than through a direct measure of the spatial complexity of the network itself. Our measurements provide quantitative evidence that the information complexity, as detected using lossless compression, reaches a maximum at the quantum critical point. \textcolor{black}{This implies that unlike Ref.~\cite{chen2025zipping}, where the ground-state transitions were associated with peaks in derivatives of CID with respect to a tuning parameter, no numerical differentiation is needed to locate the transition here.} \textcolor{black}{Although we do not have an analytical derivation for this maximum, to gain intuition, in SI we construct a toy model where the intensity of an
  image at a given pixel is obtained by multiplying a localization  envelope with a random intensity texture. One can analytically show that in this toy model,  the entropy of the image increases with the localization length, providing an intuitive
    connection between delocalization and increased image complexity. }\\

\textit{\textbf{Validation in a disordered Hofstadter model.}}
To test whether the experimental CID maximum reflects quantum Hall criticality rather than a peculiarity of one image sequence, we study a minimal model of the quantum Hall transition.
We consider the disordered Hofstadter model on a torus,
\begin{equation}
H = \sum_{\bm{r}} \mu_{\bm{r}} c^{\dagger}_{\bm{r}} c_{\bm{r}} + t \sum_{\langle \bm{r}, \bm{r'}\rangle}  \left(e^{i \phi_{\bm{r r'}}}c^{\dagger}_{\bm{r}} c_{\bm{r'}} + h.c.\right)
\end{equation}

where the phases $\phi_{\bm{r r'}}$ are chosen so that the flux through a unit cell is constant and equals $2 \pi p/q$ with $p,q$ coprime; $\mu_r$ is random onsite potential distributed uniformly between $-W/2$ and $W/2$, and $t$ is the nearest-neighbor hopping amplitude.
For a fixed disorder realization, we compute real-space LDOS maps on an $L\times L$ torus and quantify their spatial complexity using the same compression-based entropy-rate estimator. Results shown here use
$p/q = 1/10$, $W/t = 0.5$ and energy broadening $\eta/t \approx 0.02$,
which is roughly two orders of magnitude larger than the mean single-particle level spacing.

The simulated CID exhibits pronounced maxima as a function of eigenenergy (see Fig.~\ref{fig:CID}D, which shows one  disorder realization of the Hamiltonian).
Independent parameters, including the correlation length, inverse participation ratio (IPR), and average LDOS, identify the quantum Hall plateau transitions (Fig.~\ref{fig:CID}C), and the CID maxima coincide closely with these transition energies.
This supports our experimental observation that mapping spatial complexity, as quantified via the CID, can be used to locate the transition.
Representative LDOS maps show increasingly intricate spatial structure as the energy approaches \textcolor{black}{the critical point} $E_c$, providing a direct visual counterpart to the growth in CID (Fig.~\ref{fig:CID}E).
\textcolor{black}{Because the simulated images contain substantially more pixels than the experimental images, we explicitly test the effect of finite MIM spatial resolution by progressively coarse-graining the simulated LDOS landscapes. Although this changes the absolute CID, the characteristic peak near the quantum critical point remains robust as the effective resolution approaches that of the experiment (see SI; Fig. \ref{fig:CID_coarse_grained}).
This comparison indicates that a larger field-of-view-to-resolution ratio improves quantitative convergence of CID, but is not required to resolve
the quantum-critical complexity maximum studied here.}

The shape of the CID curve itself admits a simple quantitative description: modeling the measured CID as a spectrally broadened average of a localized baseline and a critical contribution, weighted by the fraction of critical spectral weight within the energy-broadening window, reproduces the observed energy dependence (see SI, Fig.~\ref{fig:functional_form}). This result supports two claims. First, it reinforces the interpretation of the experimental CID peak as a signature of the quantum Hall transition rather than an artifact of a particular device or measurement. Second, it demonstrates that a transition can be located using compression of a sequence of sufficiently large images, without requiring ensemble averaging over disorder realizations or a full finite-size-scaling analysis.
\textcolor{black}{Thus, each CID value is defined from an individual spatial configuration, while locating a transition requires a sequence of images as the tuning parameter is varied. This single-configuration character is distinct from requiring an ensemble of statistically equivalent realizations at each point.}

It should be noted that the CID identifies the \textit{location} of the transition, but extracting critical exponents or otherwise characterizing the universality class
would require systematic finite-size analysis; our claim here is limited to locating the transition, for which such scaling is not required in the present model.
As we will discuss at the end of the paper, CID also offers new insights that go beyond multifractal analysis.\\

\begin{figure*}[h!tbp]
    \includegraphics[width=0.9\textwidth]{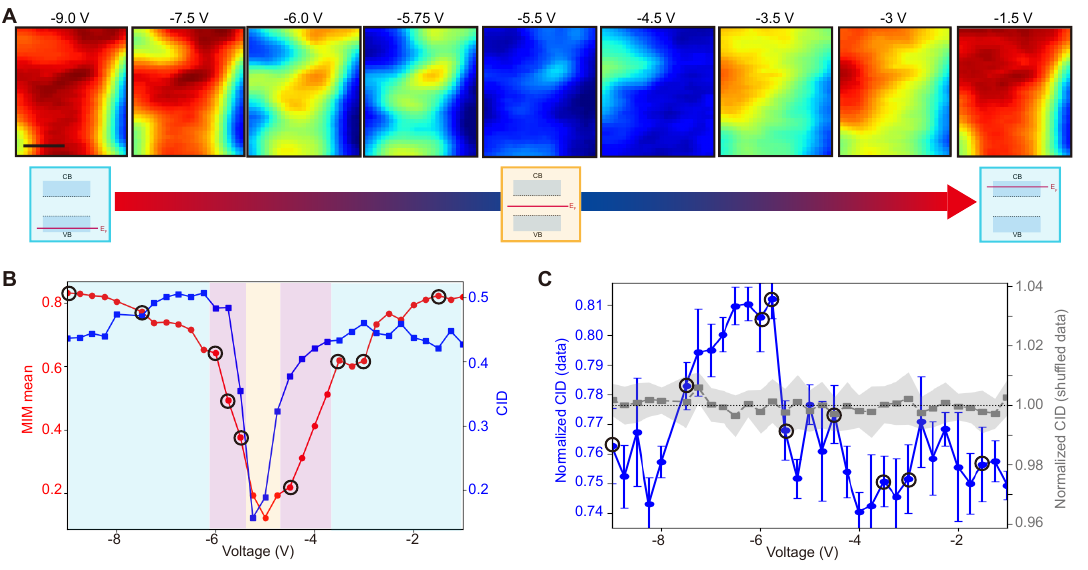}
    \caption{\label{fig:v=0 CID}
    \textbf{Mapping the real-space complexity of a trivial band insulator: the gap at charge neutrality in a MoSe$_2$/WS$_2$ moiré heterobilayer.}
    \textbf{(A)} Series of representative MIM-Im spatial maps across a trivial gapped state,
    featuring a spatially-uniform conductive phase ($-9$ V), a mixed-phase crossover ($-7.5$ to $-5.75$ V), and a nearly featureless insulating state deep within the gap ($-5.5$ V).
    Scale bar: \SI{1}{\mu m}.
    \textbf{(B)} Gate dependence of the spatially averaged MIM signal (MIM mean, red, left axis) and CID (blue, right axis) extracted from the series of MIM images. Circled dots correspond to the representative images in panel (A). \textbf{(C)} Normalized CID curve, calculated by dividing the CID of the raw data by the CID of randomly shuffled data.}
\end{figure*}

\textbf{II. Information complexity of a fractional quantum Hall liquid.}
Having established CID as a sensitive probe of integer
quantum Hall criticality both experimentally and theoretically, we next examine interaction-driven states in the fractional quantum Hall (FQH) regime.
Gate sweeps at $B = \SI{8.5}{T} $ resolve pronounced incompressible features at the fractional fillings $\nu=1/3$ and $2/3$ (Fig.~\ref{fig:Simulation}A). Despite their interaction-driven origin, plateau transitions between Abelian fractional quantum Hall states have been argued to belong to the same universality class as the integer transitions~\cite{JainKivelsonTrivedi1990,KivelsonLeeZhang1992,KumarNosovRaghu2022}. Within the composite-fermion picture, a fractional transition maps onto an integer transition of composite fermions in an effective magnetic field, motivating this expectation~\cite{Jain1989,JainKivelsonTrivedi1990}. Experimental evidence for such a superuniversality is mixed~\cite{Madathil2023,Kaur2024}.

The CID extracted from spatially resolved images across the range $0 < \nu < 1$ follows this behavior, and Fig.~\ref{fig:Simulation}B plots it together with the spatially averaged MIM-Im signal from the same image series so that the two can be compared directly. The CID peaks at the FQH transitions and decreases as the system enters each incompressible state. This suppression is weaker at fractional than at integer fillings, consistent with their smaller many-body gaps and the larger residual electronic compressibility in the presence of disorder. The real-space snapshots (Fig.~\ref{fig:Simulation}C) mirror this evolution, showing textured conductive networks between the fractional states that become more uniform upon entering the gapped phases (see Fig.~\ref{fig:Supp_Stripe} for the $\nu = 0 \to 1/3$ transition at a second location).

This dataset also illustrates a key practical feature of the diagnostic: sensitivity beyond the averaged signal.
As $\nu=1$ is approached, the CID decreases sharply near $V_g=-1.1$ V, before the spatially averaged MIM response starts to decrease (Fig.~\ref{fig:Simulation}B). The two quantities capture distinct aspects of the electronic landscape: the spatial texture can become inhomogeneous while its average response evolves more slowly; here the texture responds first.
The qualitatively similar behavior observed in the integer and fractional cases shows that CID remains sensitive to the spatial reorganization at the transition even when the gap is many-body in origin.

The results above apply to samples clean enough that a single, global transition exists within the field of view. In strongly disordered samples, however, this assumption breaks down and the spatial character of the CID diagnostic becomes essential.
To visualize this inhomogeneous behavior, we extend the CID analysis to a sliding-window implementation that produces a spatially-resolved ``entropy map'', $\mathrm{CID}(x,y)$. For an $m \times n$ MIM image, an $l \times l$ window (l=10) is translated across the frame, and each sub-image is processed using the same compression-based procedure described above, yielding an $(m-l+1)\times(n-l+1)$ spatial map of $\mathrm{CID}(x,y)$.
Regions containing mixed or strongly varying local responses exhibit elevated CID, whereas spatially uniform regions produce a lower CID.
\textcolor{black}{
Because each local CID value is calculated from
an identically sized window, finite-size compression overhead contributes approximately the same baseline throughout the map; we therefore interpret relative spatial variations rather than the absolute CID value.}

Applied near $\nu$ = 1/3 filling (Fig.~\ref{fig:Simulation}D-E), this analysis shows that the fractional state does not form uniformly across the field of view. Regions which are both incompressible and low in CID correspond to spatially coherent neighborhoods in which most nearby pixels exhibit the fractional-state response. By contrast, regions of elevated CID contain a more heterogeneous mixture of
incompressible and compressible
responses, indicating that the fractional state is spatially fragmented and
nonuniform. Therefore CID(x,y) characterizes the
local continuity of the fractional state.
Together, spatially resolved CID analysis captures the real-space disorder landscape of a fractional, interaction-driven incompressible state, providing information that is inaccessible to a global CID curve or spatially-averaged observable.
\textcolor{black}{More generally, this local implementation allows the CID to be used as an exploratory tool for identifying regions of anomalous spatial complexity or phase coexistence, even when the microscopic origin of those regions is not known in advance.}\\

\textbf{III. A control case: Complexity mapping of a trivial band insulator.}
The quantum Hall results raise an important control question: does the CID simply become large generically whenever a system crosses between a metallic and an insulating state, or does the pronounced maximum reflect the particular spatial reorganization associated with quantum Hall criticality?  To distinguish between these possibilities, we examine the $\nu = 0$ gap of an angle-aligned MoSe$_2$/WS$_2$ moiré heterobilayer, where the insulating behavior originates from a trivial single-particle band gap at charge neutrality (Fig.~\ref{fig:v=0 CID}). Tuning the Fermi energy through this gap produces a sharp conducting–insulating crossover without a QH localization transition, providing a control that separates the generic consequences of a gap opening from the distinctive spatial organization of the quantum Hall critical regime.

Within the trivial gapped state, the electronic landscape of MoSe$_2$/WS$_2$ is nearly featureless and the CID drops to a minimum (Fig.~\ref{fig:v=0 CID}A, $-5.5$ V). As the Fermi energy approaches the band edge (Fig.~\ref{fig:v=0 CID}A, $-7.5$ V $ \to -5.75$ V), charge disorder generates coexisting conductive and insulating puddles, causing the CID to rise before the spatially-averaged MIM response has fully evolved (Fig.~\ref{fig:v=0 CID}B).

\textcolor{black}{
Importantly, this crossover passes through a strongly inhomogeneous regime of coexisting conductive and insulating puddles. Yet the ratio of the measured CID to its shuffled counterpart ($CID_\textrm{data}/CID_\textrm{rand}$) does not exhibit the pronounced enhancement observed at the QH and FQH transitions (see Fig.~\ref{fig:Supp_MIM_shuffled}, Fig.~\ref{fig:v=0 CID}C). Together with the classical site-percolation benchmark, in which the CID maximum is displaced from the percolation threshold (Fig.~\ref{fig:CID_sitepercolation}), this shows that neither generic puddle formation nor geometric connectivity alone is sufficient to reproduce the quantum Hall signature. This suggests that the QH CID enhancement reflects the distinctive spatial organization of the quantum critical landscape rather than a generic consequence of gap closure or an inhomogeneous conducting–insulating crossover.}\\

\textcolor{black}{\textbf{IV. Portability across imaging modalities.} To establish the generality of entropy imaging, we applied the same compression-based analysis to independently acquired microscopy datasets spanning distinct contrast mechanisms and physical systems. In ultrafast dark-field electron microscopy of 1T-TaS$_2$, CID rises as photoinduced domains fragment and then decreases as the domain pattern coarsens; configurations with similar transformed-area fractions can nevertheless exhibit different CID values, demonstrating sensitivity to domain geometry rather than phase fraction alone (Fig.~\ref{fig:TaS2_CDW}). In X-ray magnetic circular dichroism photoemission electron microscopy (XMCD-PEEM) of FeRh, comparing the measured CID with that of randomly shuffled images distinguishes changes in magnetic domain geometry from changes in the distribution of image intensities (Fig.~\ref{fig:AFFM}).
Additional datasets show that this sensitivity to evolving geometric organization extends across other platforms:
in scanning transmission X-ray microscopy of skyrmions, CID captures field- and current-driven reorganization of real-space magnetic textures (Fig.~\ref{fig:CID_skyrmion}), and in phase-contrast microscopy of bacterial biofilms, it tracks wrinkle coarsening, arrest, and refinement (Fig.~\ref{fig:CID_biofilm}).
Together, these results show that the same compression-based framework can quantify evolving spatial organization across imaging modalities while retaining an experiment-specific physical interpretation.}\\

\textbf{Discussion.}
Entropy imaging addresses a question that conventional microscopy analysis does not directly answer: how much irreducible spatial information is contained in a measured pattern?
By applying ideas from algorithmic information theory to real-space images, CID converts spatial predictability, repetition, and complexity into an experimentally accessible observable.
\textcolor{black}{The CID can be extracted directly from a sequence of images acquired while varying temperature, field, density, pressure, or another parameter, without assuming beforehand where a phase transition occurs. Features in the CID can then identify regimes of unusual spatial reorganization, while assigning those features to a particular phase or microscopic mechanism requires complementary physical information.}

\textcolor{black}{
This distinction also highlights the complementary roles of information-theoretic probes of quantum matter. Broadly, one can consider three classes of such probes. The first includes quantities that characterize universal aspects of many-body entanglement, such as topological entanglement entropy~\cite{Kitaev06_1,levin2006detecting,satzinger2021realizing,iqbal2024topological,xu2024nonabelian,shen2026realization}. These generally require access to highly nonlocal, many-body information encoded in a state, making them difficult to measure in conventional solid-state experiments. A second class includes quantities such as quantum Fisher information, which can witness multipartite entanglement through experimentally accessible, low-order quantum correlation functions, such as frequency-resolved dynamical response functions~\cite{mazza2026quantum,hauke2016measuring,fang2025amplified,pezze2017multipartite}. Finally, CID offers a third kind of probe that extracts high-order spatial correlations and statistical dependencies from maps of experimentally accessible low-order quantum observables, such as the local density of states. It thereby probes the collective spatial organization of quantum matter without requiring direct measurements of high-order quantum correlators.}

A natural question is what CID measures beyond established image-based diagnostics. Multifractal analysis is particularly relevant near localization transitions~\cite{mandelbrot1979fractals,hentschel1983infinite,halsey1986fractal,huckestein1995scaling,evers2008anderson,bhaseen2000towards}, and it is the natural benchmark against which any new image metric should be compared.
Given a normalized LDOS $\rho(E,r)$, the system is partitioned into boxes of linear size $\ell$, and  box weights are defined as $\rho_B(\ell,E) = \sum_{r \in B} \rho(r,E)$.
From these weights one constructs partition sums $Z_q(\ell,E) = \sum_B \rho^q_B(\ell,E)$, which in turn define the multifractal exponents $\tau(q)$ at the critical energy $E= E_c$  via $Z_q(\ell, E= E_c) =  (\ell/L)^{\tau(q)}$, where $L$ is the total system size.
Because these exponents are built from the box weights alone, they encode how much intensity sits in boxes of a given size (i.e., the statistical distribution of intensities across scales) but carry no information about \textit{where} those boxes are located relative to one another. Two images can therefore share an identical multifractal spectrum while looking completely different: one a single connected network, the other a scrambled dust of the same intensities.

The CID captures exactly the difference that multifractal analysis misses, and we demonstrate this by explicit construction. Given any image, we build a surrogate ``multifractal decoy'' that has identical partition sums $Z_q(\ell,E)$ at every dyadic scale but a deliberately scrambled spatial texture. The construction is a recursive quadrant shuffle, assuming only that the box sizes $\ell$ are powers of two: we divide the image into four equal quadrants and randomly permute them, then apply the same operation recursively within each quadrant, repeatedly subdividing each block into four smaller blocks and permuting them. Because each step only rearranges the four children of a parent block and never moves intensity across parent blocks, every $Z_q(\ell,E)$, and hence every multifractal exponent, is left \textit{exactly} invariant. Yet the spatial texture (connectivity, the orientation of filaments, the long-range arrangement) is strongly altered, and the CID changes accordingly (see Fig.~\ref{fig:CID_shuffle}). Indeed, by letting the shuffle depend on energy $E$, one can even move the CID maximum away from $E_c$ (See SI for details). This establishes constructively that CID is sensitive to a class of information—the relative spatial arrangement of intensity—that no multifractal observable can access, and this is precisely the class of information in which a percolating critical network differs from a scrambled one.

The CID is also generically distinct from correlation length. As a trivial example, consider a chemical potential tuned transition between an insulator and a metal in a translationally invariant system, where the chemical potential lies within the band gap in the insulating phase and touches the bottom of the conduction band as one enters the metallic phase. The physical correlation length, defined via, say, equal-time density-density correlation functions is finite in the insulating phase while it diverges as one enters the metallic phase. However, due to translational invariance, the LDOS will be spatially uniform, and therefore CID will also be insensitive to the chemical potential and hence to the metal to insulator transition. As another example, in classical site percolation on a square lattice, the correlation length $\xi$ extracted with the same second-moment estimator diverges at the known critical threshold $p_c \approx 0.5927$, with a fitted exponent $\nu \approx 1.33$, consistent with two-dimensional percolation (Fig.~\ref{fig:CID_sitepercolation}; see SI). The CID maximum, however, occurs measurably before the correlation-length divergence. The CID is therefore not simply a monotonic repackaging of $\xi$. Their coincidence at the quantum Hall transition, in both experiment and the Hofstadter model, is a nontrivial empirical property of the quantum critical landscape.

Intertwining data compression with microscopy creates an unexpected bridge between foundational ideas in information theory and quantum matter. While lossless compression algorithms are agnostic to imaging modality, the physical interpretation of their output is not and should therefore account carefully for experiment-specific details, including spatial resolution, noise, and other measurement constraints.
For example,
it should be noted that the data inherit the instrument transfer function together with image size, discretization and compression overhead.
\textcolor{black}{In particular, the finite number of pixels limits how closely the CID approaches its asymptotic entropy rate value. Increasing either the field of view or the spatial resolution improves this convergence; consequently, comparisons of absolute CID across datasets with different dimensions or resolutions should be done with care. In this work, the transition signature relies only on relative changes in the information complexity across a series of images acquired and analyzed under consistent conditions.}
Additionally, random noise can be algorithmically incompressible without being physically interesting, so absolute CID values should not be compared indiscriminately across unrelated measurements.

Portability is the technique's most immediately practical feature.
Because the analysis operates on the pattern itself regardless of imaging modality, it can be applied across scanning probe, electron, x-ray, optical, magnetic, and other imaging platforms (see Supplementary Information).
This creates opportunities to quantify emergent spatial organization not only in quantum materials, but across other systems in physics, materials science, chemistry, biology, and beyond.
It can also be applied retroactively to existing datasets at essentially no cost.

\begin{acknowledgments}
The authors thank Richard Averitt and Daniel Parker for insightful discussions, and Kohtaro Yamakawa and James Analytis for providing the Eu$_5$Sn$_2$As$_6$ sample.
M.T.A and A.F.Y were primarily supported by NSF through the UC San Diego Materials Research Science and Engineering Center (UCSD MRSEC), grant number DMR-2614051. T.G. is supported by the National Science Foundation under Grant No. DMR-2521369. The TMD heterostructure device fabrication is supported by the Office of Basic Energy Sciences, Materials Sciences and Engineering Division, of the U.S. Department of Energy under Contract DE-AC02- 05-CH11231, within the van der Waals Heterostructures Program (KCWF16). Graphene device fabrication was performed, in part, at the San Diego Nanotechnology Infrastructure (SDNI) of UCSD, a member of the National Nanotechnology Coordinated Infrastructure, which is supported by the National Science Foundation (Grant ECCS-2025752).
\end{acknowledgments}

\bibliography{CID_bib}

\newpage

\begin{widetext}

\section*{Supplementary Information}

\startcontents
\renewcommand{\thefigure}{S\arabic{figure}}
\setcounter{figure}{0}
\renewcommand{\thesection}{\arabic{section}}
\renewcommand{\thesubsection}{\thesection.\arabic{subsection}}
\makeatletter\renewcommand{\p@subsection}{}\makeatother

\FloatBarrier
\section{CID analysis of quantum Hall transitions}
\FloatBarrier
\subsection{Experimental Methods}

The single-gated graphene device was fabricated using a standard dry transfer method. Both hBN and graphite were mechanically exfoliated onto silicon (Si)/silicon dioxide (SiO$_2$) substrates, and suitable flakes were identified using an optical microscope. Graphene and graphite flakes went through thermal annealing in an H$_2$/N$_2$ environment at 350$^{\circ}$C for 6 hours. A PDMS stamp covered with a thin polycarbonate film was used to pick up flakes from top to bottom in the order thin hBN - MLG - thick hBN - graphite. Top and bottom hBN are \SI{10}{nm} and \SI{90}{nm} separately. The bottom hBN was chosen to be thick enough to avoid the saturation of MIM signal from the back-gate graphite. The entire heterostructure was deposited onto a substrate with pre-patterned contacts (Ti/Pt with thickness of 2nm/10nm) at 180$^{\circ}$C. After dissolving the polycarbonate film in chloroform, mechanical AFM cleaning was performed on the surface of the heterostructure \cite{goossens2012AFMClean}.

The MoSe$_2$/WS$_2$ heterobilayer was fabricated using a dry-transfer method based on polyethylene terephthalate glycol (PETG) stamps. Monolayer MoSe$_2$, monolayer WS$_2$, few-layer graphite and hBN flakes are mechanically exfoliated from bulk crystals onto Si substrates with a 90-nm-thick SiO$_2$ layer. A 0.5 mm-thick clear PETG stamp is employed to pick up the flakes at 65-75°C sequentially, in the order of top graphite gate (trilayer), top hBN (26 nm), monolayer WS$_2$, monolayer MoSe$_2$, graphite contact (few-layer), bottom hBN (50 nm). The whole stack is then released onto a pre-patterned 60nm Au electrode with 5nm Cr adhesion layer on a 90 nm SiO$_2$/Si substrate (intrinsic) at 95-100°C. The PETG is then dissolved in chloroform at room temperature overnight. Electrodes (Cr/Au) are defined using photolithography (Durham Magneto Optics, MicroWriter) and evaporated using electron-beam evaporation (Angstrom Engineering).

On the instrument side, the setup uses the same components as reported in \cite{cao2023millikelvin, Cao2025layer}. During measurements, a microwave excitation of \SI{-35}{dBm} at \SI{4.22}{GHz} was delivered to the tip apex, and the reflected microwave signal was amplified and demodulated to extract the tip-sample admittance.

\FloatBarrier
\subsection{Serialization of non-power-of-two images}\label{sec:serialization}

In this section we briefly discuss how the Hilbert curve is defined in a canvas where the pixel number is not $2^{n}$. The Hilbert construction is defined on a $2^{n}\times 2^{n}$ domain,
whereas each MIM frame, for example, is $40\times40$ pixels. We therefore embed every
frame in the upper-left corner of a $64\times64$ grid, leave the
remaining entries undefined, generate the curve on the full grid, and
delete the undefined entries from the resulting string. Because the
$40\times40$ region decomposes into complete dyadic blocks, this
embedding preserves spatial locality almost exactly: all but one of the
$1599$ consecutive pairs in the reduced string are nearest neighbours in
the image.

\FloatBarrier
\textcolor{black}{\subsection{Convergence of CID to the Shannon entropy rate: exactly solvable benchmarks}
For a random field, the Shannon entropy rate $h$ is the irreducible information per site once all spatial correlations are accounted for. By the source-coding theorem, no lossless code can on average use fewer than $h$ bits per site, and universal Lempel--Ziv compressors approach this bound as the system size increases. CID is therefore expected to approach $h$ from above for large images. We test this on two models whose entropy rate is known exactly: uncorrelated site percolation and the two-dimensional Ising model. Because the aim here is to test convergence under near-ideal conditions, configurations are stored with one bit per site, serialized along a Hilbert curve, and compressed with LZMA (Lempel--Ziv with adaptive range coding). With this encoding, CID is directly the number of compressed bits per site and can be compared with the entropy without rescaling.}

\begin{figure}[htbp]
    \includegraphics[width=0.85\textwidth]{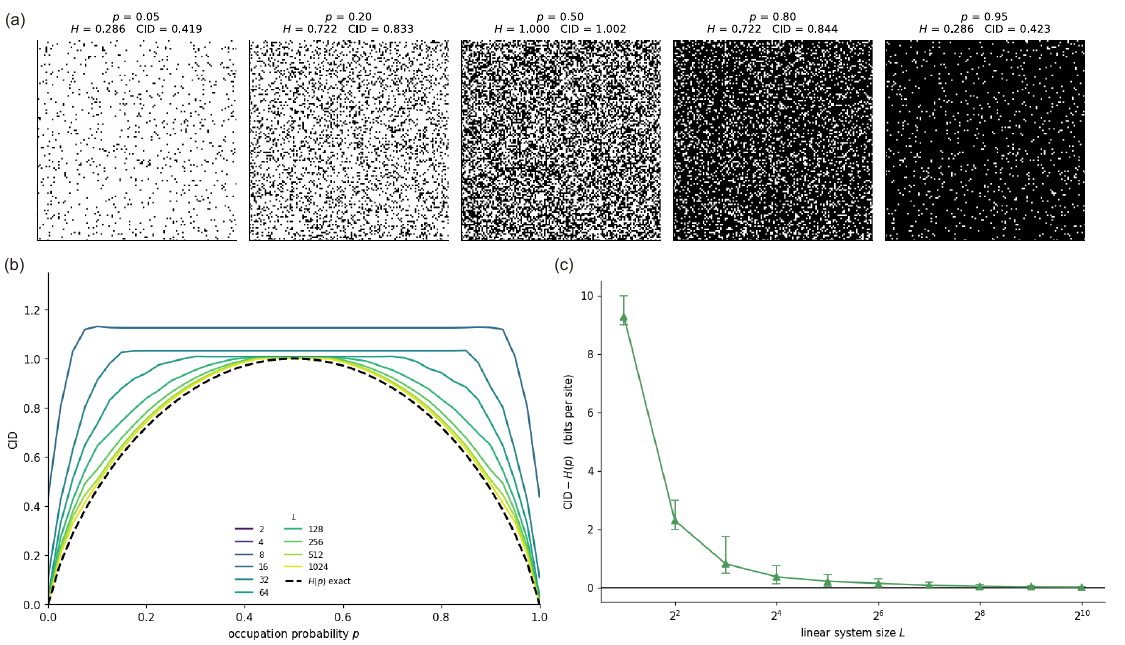}
\caption{\label{fig:SI_sitepercolation}\textbf{Convergence of CID to the Shannon entropy: site percolation model.}
(a) Example configurations ($L=128$) at the indicated occupation probabilities $p$.
(b) CID versus $p$ for $L\times L$ lattices, $L=2$--$1024$ (mean of 20 realizations). Black dashed line: exact $H(p)$.
(c) Excess CID$\,-\,H(p)$ versus $L$, error bars indicate the spread of the excess across $p$.
}
\end{figure}

\textcolor{black}{\subsubsection{Site percolation}
In site percolation each site is occupied independently with probability $p$ (Fig.~\ref{fig:SI_sitepercolation}a), so the entropy rate is exactly the binary entropy $H(p)=-p\log_2p-(1-p)\log_2(1-p)$. We generated $L\times L$ configurations for $L=2$--$1024$. For the smallest images where all pixels are quickly flipped, CID is dominated by fixed compressor overhead and is nearly independent of $p$. As $L$ increases, CID approaches $H(p)$ from above over the whole range of $p$ (Fig.~\ref{fig:SI_sitepercolation}b). Averaged over $p$, the excess CID$\,-\,H(p)$ decreases from about 9 bits per site at $L=2$ to around 0.02 bits per site at $L=1024$ (Fig.~\ref{fig:SI_sitepercolation}c). The maximum of CID lies at $p=1/2$, set by the occupation statistics rather than by the percolation threshold.}

\begin{figure}[htbp]
    \includegraphics[width=0.9\textwidth]{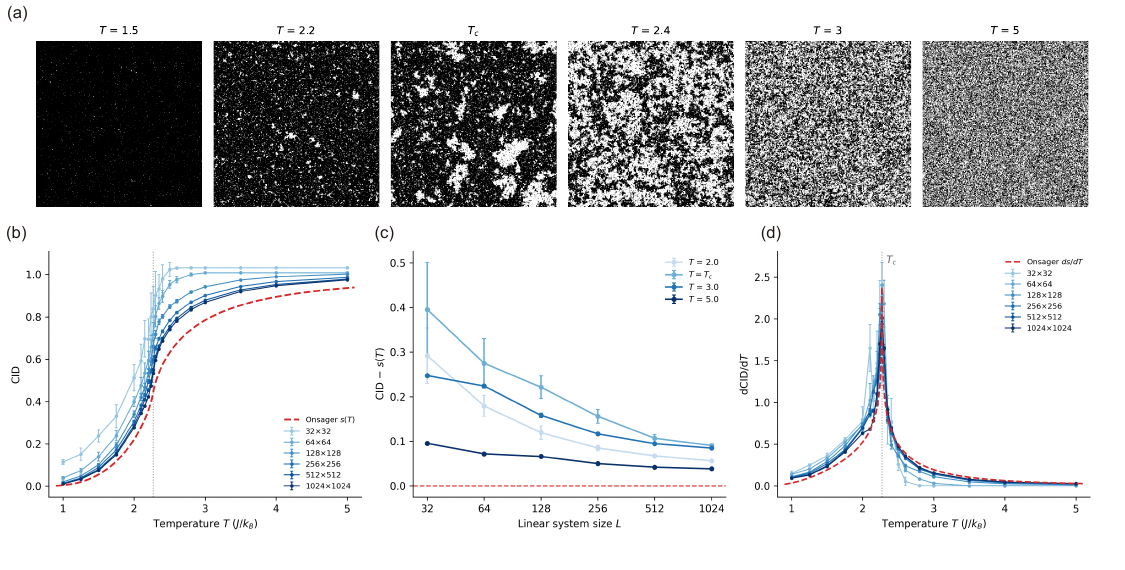}
\caption{\label{fig:SI_ising}\textbf{Convergence of CID to the Shannon entropy: two-dimensional Ising model.}
(a) Equilibrium configurations for $L=512$ at the temperatures indicated.
(a) Equilibrium configurations for $L=512$ at different temperatures.
(b) CID versus temperature $T$ for $L=32$--$1024$. Points are means over 20 configurations, with error bars of $\pm1$ s.d. Red dashed line: exact Onsager entropy per spin $s(T)$.
(c) CID$\,-\,s(T)$ versus $L$ at $T=2.0$, $T_c$, 3.0 and 5.0.
(d) Temperature derivative dCID/d$T$, obtained by numerical differentiation of the mean curves; error bars are bootstrap estimates. Red dashed line: exact $ds/dT=c/(T\ln2)$, where $c$ is the specific heat per spin.
Dotted lines in (b) and (d) mark $T_c=2/\ln(1+\sqrt2)\approx2.269$.
}
\end{figure}

\textcolor{black}{\subsubsection{Two-dimensional Ising model}
To test a system with spatial correlations and a genuine phase transition, we sampled equilibrium configurations of the square-lattice Ising model with periodic boundaries using the Swendsen--Wang cluster algorithm for $L=32$--$1024$. For equilibrium configurations, the Shannon entropy rate equals the thermodynamic entropy per spin, which is known exactly from Onsager's solution, $s(T)=\ln Z/N+\beta u$ (converted to bits). The sampled energies agree with Onsager's exact energy to within 0.2\%.
CID follows $s(T)$ from above at all temperatures and approaches it steadily as $L$ increases (Fig.~\ref{fig:SI_ising}b,c). The excess CID$\,-\,s(T)$ is largest near $T_c$, where long-range correlations require larger images for the compressor to capture fully. Even there it decreases from 0.40 bits at $L=32$ to 0.09 bits at $L=1024$, and at $T=5$ it is below 0.04 bits. CID itself has no maximum at the transition: like the entropy, it increases monotonically with $T$, most steeply near $T_c$. The transition appears instead as a sharp peak in dCID/d$T$ at $T_c$ for all $L\geq64$, closely following the exact $ds/dT=c/(T\ln2)$, where $c$ is the specific heat (Fig.~\ref{fig:SI_ising}d). Thus, for a conventional thermal transition, the temperature derivative of CID reproduces the thermodynamic signature of the transition directly from real-space configurations. This contrasts with the quantum Hall transitions in the main text, where CID itself is maximal at the transition.}

\FloatBarrier
\subsection{A toy model relating localization length to entropy of an image}\label{sec:toymodelcid}
In the IQH problem considered in the main text, the LDOS at a given energy receives contributions from single-particle eigenstates within the energy-resolution window. Their localization length grows on approaching the plateau
  transition, while their intensity profiles also exhibit disorder-induced spatial fluctuations. Motivated by the connection between lossless compression and Shannon entropy, we consider a minimal model for a single localized intensity profile with random spatial modulation. Different realizations of this modulation define an ensemble of
  images. We show analytically that the joint entropy of the continuous image intensities increases with the localization length when the probability distribution of the random modulation is independent of that length.

We consider a two-dimensional lattice with $L$ sites along each side and
$N=L^2$ pixels.  We label the pixels by $i=1,\ldots,N$, with pixel $i$ at
position $\mathbf r_i=(x_i,y_i)$.  Choose a center $\mathbf r_0$ and let
$d_i=|\mathbf r_i-\mathbf r_0|$ be the distance to that center.  We assign
the intensity
\begin{equation}
p_i(\xi)=
	\frac{\overbrace{e^{-2d_i/\xi}}^{\text{envelope}}\,\,
		\overbrace{e^{s z_i}}^{\text{random factor}}}
	{\displaystyle\sum_{j=1}^N e^{-2d_j/\xi+s z_j}}.
\label{eq:appendix-speckle}
\end{equation}
Here $\xi>0$ is the ``localization length'' associated with the envelope.  Each $z_i$ is an independent Gaussian number with mean zero and variance one, and $s>0$ sets the strength of the brightness fluctuations. The denominator
makes $\sum_i p_i=1$. We assume that $s$ is independent of $\xi$.

Let $f_\xi(p_1,\ldots,p_{N-1})$ be the joint probability density of the
pixel values.  In particular, the probability that every one
of these values falls in a specified small interval of width $\Delta$
is approximately $f_\xi\Delta^{N-1}$.
The continuous analogue of the Shannon entropy of this ensemble of images is the differential entropy
\begin{equation}
h_{\rm image}(\xi)
=-\int f_\xi\log_2 f_\xi\,dp_1\cdots dp_{N-1} =-\E[\log_2 f_\xi].
\label{eq:appendix-differential-definition}
\end{equation}
The integral is over positive values with
$\sum_{i=1}^{N-1}p_i<1$.  Note that the normalization determines \(p_N=1-\sum_{i<N}p_i\), and therefore the first $N-1$ intensities specify the entire image. Above, \(\E\) denotes the averaging over the Gaussian random variables.

To calculate this, it is easier to
work with the log ratios
\begin{equation}
u_i=\ln\frac{p_i}{p_N}
=s(z_i-z_N)-\frac{2(d_i-d_N)}{\xi},\qquad i=1,\ldots,N-1,
\label{eq:appendix-log-ratios}
\end{equation}
which is now a linear function of variables whose distribution is
Gaussian.  Let's denote the joint probability density of $u_i$ by $q_\xi$.  Since the probability is
preserved when changing variables,
$f_\xi\,dp_1\cdots dp_{N-1}=q_\xi\,du_1\cdots du_{N-1}$.
If $J$ is the absolute Jacobian determinant for changing from the
$u_i$ to the $p_i$, then $f_\xi=q_\xi/J$.  Substituting this into Eq.\eqref{eq:appendix-differential-definition},
\begin{equation}
h_{\rm image}=-\E[\log_2(q_\xi/J)]
=-\E[\log_2 q_\xi]+\E[\log_2 J].
\label{eq:appendix-entropy-change-variables}
\end{equation}
It turns out the first term is independent of $\xi$.  Indeed, Eq.~\eqref{eq:appendix-log-ratios} has the form
\begin{equation}
u_i=\mu_i(\xi)+v_i,\,
\mu_i(\xi)=-\frac{2(d_i-d_N)}{\xi},\,v_i=s(z_i-z_N).
\label{eq:appendix-translation}
\end{equation}
The $v_i$ have the same joint distribution for every $\xi$: changing
$\xi$ only shifts the means.  Let $q_\infty$ denote their density,
which is also the density of the $u_i$ at $\xi=\infty$.  Writing
$\mathbf u$, $\boldsymbol\mu$, and $\mathbf v$ for the corresponding
vectors, $q_\xi(\mathbf u)=q_\infty(\mathbf u-\boldsymbol\mu)$.
Changing the integration variable to $\mathbf v=\mathbf u-\boldsymbol\mu$
therefore gives
\begin{equation}
\begin{aligned}
-\E[\log_2q_\xi]
&=-\int q_\infty(\mathbf u-\boldsymbol\mu)
\log_2 q_\infty(\mathbf u-\boldsymbol\mu)\,d\mathbf u\\
&=-\int q_\infty(\mathbf v)\log_2q_\infty(\mathbf v)\,d\mathbf v
\equiv C_{N,s}.
\end{aligned}
\label{eq:appendix-constant-by-translation}
\end{equation}
Here $d\mathbf u=du_1\cdots du_{N-1}$ and similarly for $d\mathbf v$.
This is clearly $\xi$ independent. Therefore, all the $\xi$-dependence of $h_{\rm image}$ is in
$\E[\log_2J]$.

To calculate $\E[\log_2 J]$, we first express the intensities in terms
  of the log ratios:
  \begin{equation}
  p_i=\frac{e^{u_i}}{1+\sum_{j<N}e^{u_j}}\quad(i<N),
  \qquad
  p_N=\frac{1}{1+\sum_{j<N}e^{u_j}}.
  \end{equation}
  Differentiating gives, for $i,j<N$,
  \begin{equation}
  \frac{\partial p_i}{\partial u_j}
  =
  \begin{cases}
  p_i(1-p_i),&i=j,\\
  -p_i p_j,&i\ne j.
  \end{cases}
  \label{eq:appendix-jacobian-entries}
  \end{equation}
  Let $\mathbf p=(p_1,\ldots,p_{N-1})^{\mathsf T}$, with $\mathbf1$
  the vector of $N-1$ ones and $\mathbb I$ the identity matrix of that
  size. Factoring $p_i$ from each row gives
  \begin{equation}
  \begin{aligned}
  J&=\left(\prod_{i<N}p_i\right)
  \det(\mathbb I-\mathbf1\mathbf p^{\mathsf T})\\
  &=\left(\prod_{i<N}p_i\right)
  \left(1-\sum_{i<N}p_i\right)
  =\prod_{i=1}^N p_i.
  \end{aligned}
  \label{eq:appendix-jacobian-determinant}
  \end{equation}
  The second equality follows because $\mathbf1\mathbf p^{\mathsf T}$
  has one nonzero eigenvalue, $\sum_{i<N}p_i$.
  Thus $\E[\log_2 J]=\sum_i\E[\log_2 p_i]$, and
  \begin{equation}
  h_{\rm image}(\xi)
  =C_{N,s}+\sum_{i=1}^N\E[\log_2 p_i(\xi)].
  \end{equation}

Our main goal is to understand if $\frac{d h_{\rm image}}{d\xi}$ has a definite sign. Writing $t=2/\xi$,
$d h_{\rm image}/dt=(\ln2)^{-1}\sum_i\E[\partial_t\ln p_i]$.
The averages are over the same Gaussian law for every $t$.
Differentiating the expression for $p_i$, Eq.\eqref{eq:appendix-speckle}, gives
\begin{equation}
\partial_t\ln p_i
=-d_i+\frac{\sum_j d_j e^{-t d_j+s z_j}}
{\sum_j e^{-t d_j+s z_j}}
=-d_i+\sum_jp_jd_j.
\end{equation}
This implies
\begin{equation}
\frac{d h_{\rm image}}{dt}
=\frac{N}{\ln2}\left(\E\!\left[\sum_i p_i d_i\right]
-\frac1N\sum_i d_i\right).
\label{eq:appendix-entropy-first-derivative}
\end{equation}
To see how this derivative changes with $t$, we use
$\partial_t p_i=p_i\partial_t\ln p_i
=p_i(-d_i+\sum_jp_jd_j)$ in the remaining expectation:
\begin{equation}
\begin{aligned}
\frac{d^2 h_{\rm image}}{dt^2}
&=\frac{N}{\ln2}\E\!\left[\sum_i d_i p_i
\left(-d_i+\sum_jp_jd_j\right)\right]\\
&=-\frac{N}{\ln2}\E\!\left[\sum_i p_i d_i^2-
\left(\sum_i p_i d_i\right)^2\right]\leq0.
\end{aligned}
\label{eq:appendix-convexity}
\end{equation}
where the inequality follows from the  fact that the object inside the $[\,\,]$ brackets is a variance.  At $t=0$ all pixels have the
same mean intensity $\E[p_i]=1/N$, and
Eq.~\eqref{eq:appendix-entropy-first-derivative} gives
$d h_{\rm image}/dt=0$.  The nonpositive second derivative implies
$d h_{\rm image}/dt\leq0$ for $t\geq0$.  Finally, $dt/d\xi=-2/\xi^2$,
so
\begin{equation}
\boxed{\frac{d h_{\rm image}}{d\xi}
	=-\frac{2}{\xi^2}\frac{d h_{\rm image}}{dt}\geq0.}
\label{eq:appendix-monotonicity}
\end{equation}
The maximum lies at $\xi\to\infty$.

The physical picture is simple: a localized image has many dim pixels
whose fluctuations occupy a small absolute intensity range.  A weaker
envelope spreads resolvable fluctuations over more pixels. This result concerns the joint entropy of the toy model considered here and provides intuition, but it is not a proof for the monotonic dependence of CID on the localization length in the IQH problem.

\FloatBarrier
\subsection{Functional form study of CID}
In the main text we stated that the energy dependence of the simulated CID
admits a simple quantitative description: the CID behaves as a spectrally
broadened average of a localized baseline and a critical contribution,
weighted by the fraction of critical spectral weight within the
energy-broadening window. Here we make this model explicit and fit it to the
CID extracted from the Hofstadter LDOS maps.
\paragraph*{Two-level ansatz.}
The LDOS at probe energy $E$ is a broadened sum over eigenstates,
$\mathrm{LDOS}(\bm{r},E)=\sum_n |\psi_n(\bm{r})|^2\, g_\eta(E-E_n)$, where
$g_\eta$ is a normalized Gaussian of width $\eta$. Suppose that each
eigenstate $n$ contributes a characteristic spatial complexity $c_n$ to the
image, so that the complexity probed at energy $E$ is the spectrally weighted
average
\begin{equation}
C(E,\eta)=\frac{\sum_n g_\eta(E-E_n)\,c_n}{\sum_n g_\eta(E-E_n)} .
\label{eq:Cdef}
\end{equation}
We adopt a two-level model,
\begin{equation}
c_n = c_{\mathrm{loc}} + \left(c_{\mathrm{crit}}-c_{\mathrm{loc}}\right)\chi_n ,
\label{eq:twolevel}
\end{equation}
where $\chi_n=1$ if state $n$ is critical (i.e., it lies within the critical
window at the center of its subband) and $\chi_n=0$ if it is localized;
$c_{\mathrm{loc}}$ and $c_{\mathrm{crit}}$ denote the complexity of the
localized floor and of the critical states, respectively. Substituting
Eq.~\eqref{eq:twolevel} into Eq.~\eqref{eq:Cdef} gives

\begin{equation}
\begin{aligned}
C(E,\eta)&=c_{\mathrm{loc}} + \left(c_{\mathrm{crit}}-c_{\mathrm{loc}}\right) P_\eta(E),\\
P_\eta(E)&=\frac{\sum_n g_\eta(E-E_n)\,\chi_n}{\sum_n g_\eta(E-E_n)} ,
\end{aligned}
\label{eq:Peta}
\end{equation}

so that $P_\eta(E)\in[0,1]$ is the fraction of critical spectral weight
within the broadening window centered at $E$. Assuming an affine relation
between the measured CID and $C$, we arrive at
\begin{equation}
\mathrm{CID}(E)=B_0 + B_1\, P_\eta(E),
\label{eq:cidmodel}
\end{equation}
where $B_0$ is the CID of the localized floor and $B_1$ is the amplitude of
the critical enhancement.

\paragraph*{Criticality profile and exact-spectrum evaluation.}
We evaluate Eq.~\eqref{eq:Peta} directly with the exact eigenvalue
spectrum $\{E_n\}$ of the similar disorder realization used to generate
the LDOS maps, so that the density of states enters the model exactly
and requires no parameterization. The only modeled quantity is the
criticality profile: each eigenstate is assigned the weight
$c_n = c(E_n)$, with
\begin{equation}
c(E') = 1 - \prod_k \left[\, 1 -
\exp\!\left(-\frac{(E'-\mu_k)^2}{2\Gamma_k^2}\right) \right],
\label{eq:cmodel}
\end{equation}
a saturating product of Gaussian windows centered at the subband
centers $\mu_k$ of the clean Harper spectrum. For isolated windows
this reduces to a single Gaussian window; where windows overlap it
saturates at one, so that $c \in [0,1]$ and hence $P_\eta(E) \le 1$
by construction, making $B_0 + B_1$ a strict upper bound on the model
CID. The window widths inherit the clean subband structure,
\begin{equation}
\Gamma_k^2=\left(w\,W_k\right)^2+\Gamma^2 ,
\label{eq:widths}
\end{equation}
where $W_k$ is the bandwidth of the $k$-th subband of the clean
Harper spectrum, $w$ is a global scale factor, and $\Gamma$ is a
uniform floor. The centers $\mu_k$ and clean bandwidths $W_k$ are
fixed by the Harper spectrum at $\alpha = p/q = 1/10$ and are not
adjusted in the fit. (In the limit of a Gaussian-broadened density
of states and non-overlapping critical windows, Eq.~\eqref{eq:Peta}
admits a closed form as a ratio of Gaussian sums; we do not use it
here.)

\paragraph*{Fit.}
The model contains four free parameters: the window floor $\Gamma$,
the window scale $w$, the baseline $B_0$, and the enhancement
amplitude $B_1$. The energy broadening is fixed to the value used to
generate this CID series, $\eta/t = 0.04$. A nonlinear least-squares fit over $|E/t| \le 3$
converges to an interior point of the parameter space,
$\Gamma/t = 0.0661 \pm 0.0010$, $w = 3.25 \pm 0.09$,
$B_0 = 0.6253 \pm 0.0003$, and $B_1 = 0.0229 \pm 0.0003$, with an
rms residual of $2.05\times10^{-3}$. The fit reproduces the peak
positions, the peak widths, and the saturated plateau at the band
center, where the overlapping critical windows of the two central
subbands drive $P_\eta \to 1$ and the CID approaches its ceiling
$B_0 + B_1$ (Fig.~\ref{fig:functional_form}). The fitted $w \approx 3$ indicates that the
energy window over which states behave critically in this finite
system is several times broader than the clean subband width.

At the center of each spectral gap, the $\eta$-broadened LDOS
switches from being dominated by states at the lower gap edge to
states at the upper edge over an energy window of order
$\eta^2/\Delta$, where $\Delta$ is the gap size. Because the
localized textures on the two edges differ in complexity, both the
measured CID and the model exhibit a small, sharp step at the gap
centers ($|E/t| \approx 1.1$, $1.9$, $2.9$; most visible near
$E/t \approx -1.9$). The model reproduces the location and sign of
these steps, though not their precise magnitude, since it assigns
each eigenstate a complexity determined only by its energy. The
remaining systematic deviation on the flanks of the peaks reflects
the simplifying assumption of Gaussian windows centered at the clean
subband energies.

\begin{figure}[htbp]
    \includegraphics[width=0.6\textwidth]{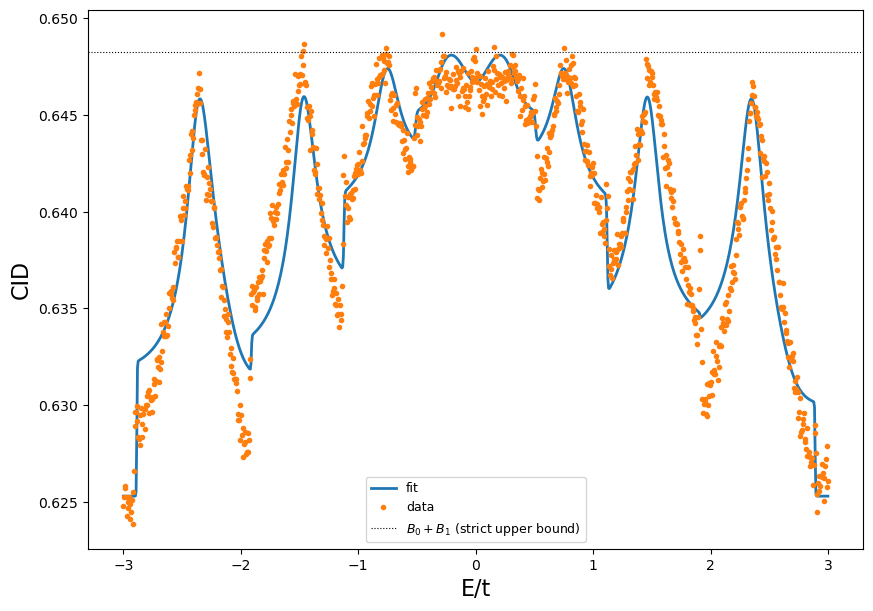}
\caption{\label{fig:functional_form} \textbf{Functional form of the CID across the
disorder-broadened Hofstadter subbands.} CID of the simulated LDOS
maps (orange points; one disorder realization, $p/q = 1/10$,
$W/t = 0.5$, $\eta/t = 0.04$) as a function of energy, together with
the four-parameter fit of Eqs.~\eqref{eq:Peta}, ~\eqref{eq:cidmodel}, and~\eqref{eq:cmodel} (blue line), evaluated with the exact eigenvalue spectrum of the same
disorder realization. The dotted line marks $B_0 + B_1$, the strict
upper bound of the model, which the data reach only at the band
center, where the critical windows of the two central subbands
overlap. Small steps at the gap centers, present in both the data
and the model, reflect the crossover between states at the lower and
upper gap edges (see text). Fitted parameters:
$\Gamma/t = 0.0661 \pm 0.0010$, $w = 3.25 \pm 0.09$,
$B_0 = 0.6253 \pm 0.0003$, $B_1 = 0.0229 \pm 0.0003$.}
\end{figure}

\FloatBarrier
\subsection{Normalized CID: separating spatial arrangement from the intensity distribution}\label{sec:normCID}

\textcolor{black}{The CID of an image depends both on the spatial arrangement of its pixel values and on their distribution: an image whose values occupy only a few levels compresses well regardless of how they are arranged. To separate the two contributions, we compare each image with surrogates that contain exactly the same pixel values placed at random positions. For each image, we randomly permute the pixel positions, compute the CID of the permuted image with the same serialization and compressor, and average over $N$ independent permutations ($N=10$) to obtain $CID_{\mathrm{rand}}$. The normalized CID is defined as $CID_{\mathrm{data}}$/$CID_{\mathrm{rand}}$. Because the permutation preserves the histogram but destroys all spatial correlations, the normalized CID largely removes the contribution of the intensity distribution; values below one indicate spatial structure that the compressor can exploit.}

We apply this procedure to the integer and fractional quantum Hall data from the main text (Fig.~\ref{fig:Supp_MIM_shuffled}), which shows that CID will depend on the geometrical complexity after normalizing it by dividing the CID of the shuffled version of the same image.
The transition will be persistent after getting rid of the histogram information. The normalized CID for the sequence of the  shuffled plot will be mostly within $1.00 \pm 0.01$ through 10 rounds of randomization.

As a comparison, we applied the same method on the simulated LDOS (Fig.~\ref{fig:Supp_LDOS_shuffled}). The result confirms that the normalized CID also removes all the histogram information and the remaining CID captures the geometric information, similar to the experimental cases. The dips and peaks also match the CID of the LDOS texture in the main text (Fig.~\ref{fig:CID}).

 \begin{figure}[h!tbp]
     \includegraphics[width=0.8\textwidth]{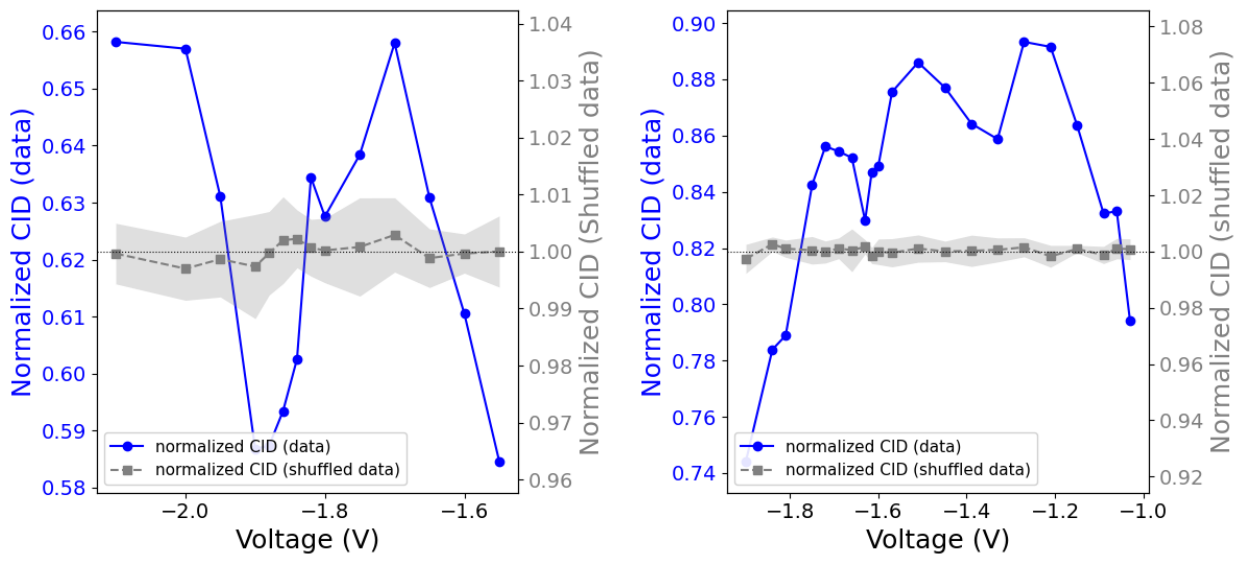}
     \caption{\label{fig:Supp_MIM_shuffled}
     \textbf{Comparison of normalized CID on data and shuffled data}
(a) Comparison of normalized CID on data and shuffled data of IQHE
(b)  Comparison of normalized CID on data and shuffled data of FQHE.
     }
 \end{figure}

  \begin{figure}[h!tbp]
     \includegraphics[width=0.5\textwidth]{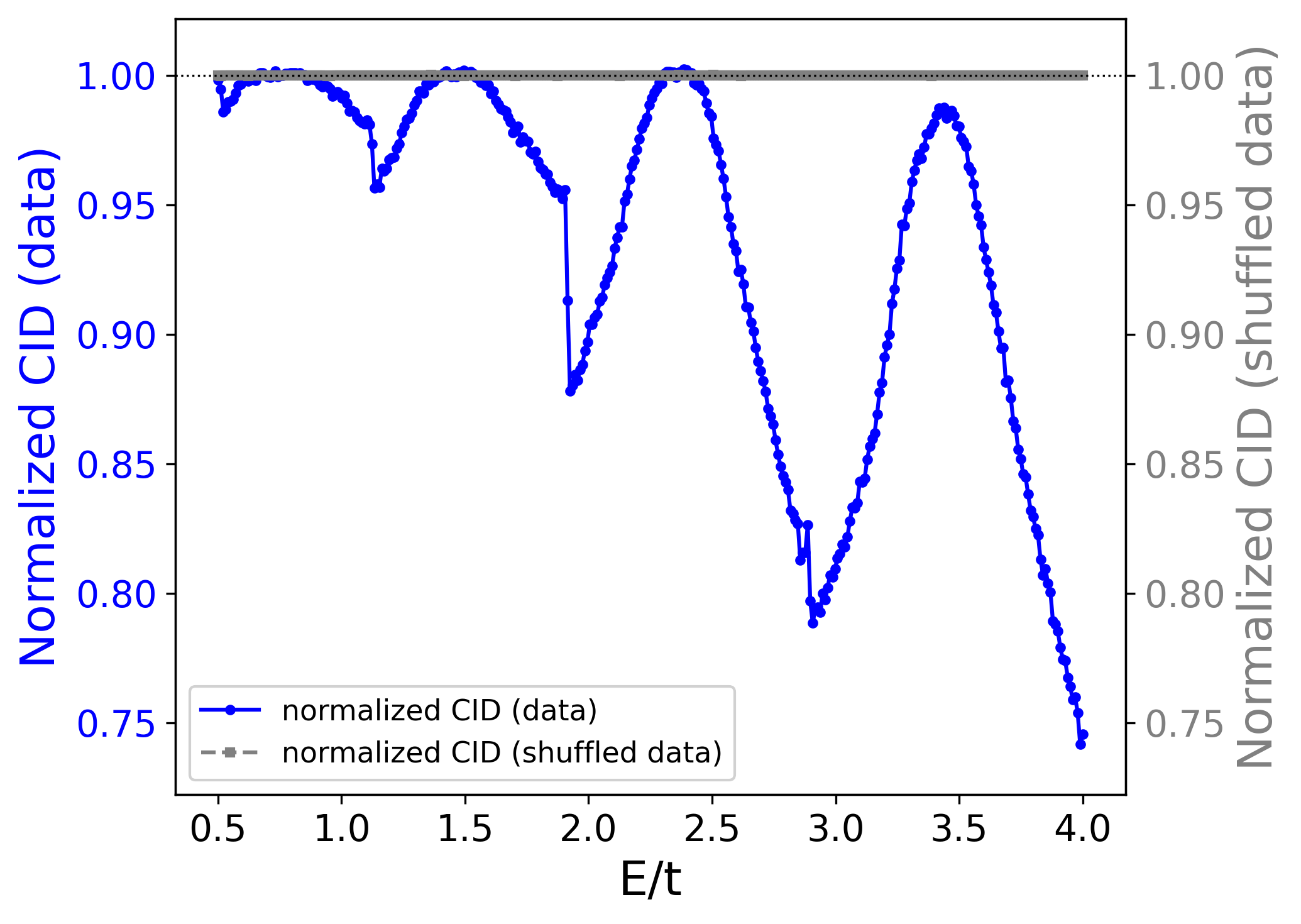}
     \caption{\label{fig:Supp_LDOS_shuffled}
     \textbf{Apply normalized CID to the simulated LDOS and shuffled LDOS}
    The same algorithm of normalized CID is applied to the simulated LDOS from the Hofstadter model.
     }
 \end{figure}

\FloatBarrier
\subsection{A shuffling procedure that changes CID but keeps multifractal exponents invariant}

 \begin{figure}[h!tbp]
     \includegraphics[width=0.7\textwidth]{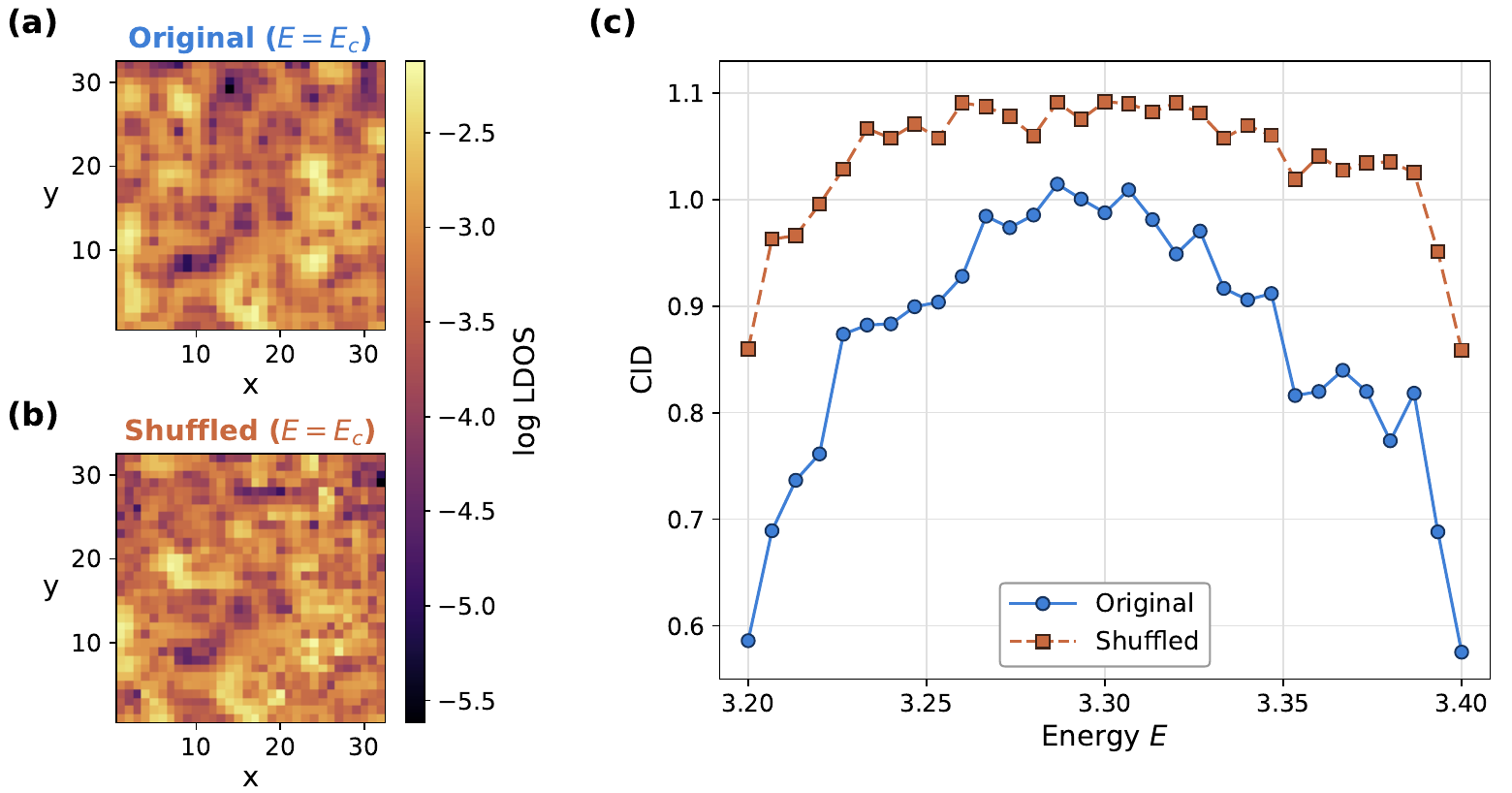}
     \caption{\label{fig:CID_shuffle}
     \textbf{Comparison between original and shuffled LDOS images demonstrating invariance of multi-fractal statistics and sensitivity of CID to spatial correlation.}
     \textbf{(a)} Original LDOS map of the Hofstadter model at $E = E_c$ (log scale).
\textbf{(b)} Quadrant-shuffled result of (a), with identical $Z_q(\ell)$ at all dyadic scales but a different spatial texture.
\textbf{(c)} CID of the original and shuffled maps versus energy $E$.
     }
 \end{figure}

In this section we discuss the idea mentioned in the main text that illustrates the difference between scale-dependent intensity statistics (as captured by the standard multifractal analysis~\cite{mandelbrot1979fractals,hentschel1983infinite,halsey1986fractal,huckestein1995scaling,evers2008anderson,bhaseen2000towards}) and the complexity of spatial texture, as captured by CID. We underline this distinction by constructing a surrogate LDOS image that preserves the box-weights used in multifractal analysis while strongly rearranging the spatial pattern. The rearrangement procedure works as follows: the input is an image with normalized, nonnegative spatial density $\rho(\bf{r})$, i.e., $\sum_{\bf{r}} \rho(\bf{r}) = 1$. We restrict to system sizes that are some power of two, i.e., $L = 2^n$ so that the system admits a natural dyadic hierarchy of partitions. At dyadic scale $\ell = 2^k$, the image is tiled by $(L/\ell)^2 =4^{n-k}$ number of non-overlapping boxes $B$, which allows one to define multifractal partition sums $Z_q(\ell) = \sum_B \rho_B(\ell)^q$ where  $\rho_B(\ell) = \sum_{\bf{r} \in B}\rho(\bf{r})$ is the total density inside the box $B$. Our goal is to modify the arrangement of these boxes so that $Z_q(\ell)$ remains invariant. This is done by a recursive ``quadrant shuffle''---starting from the full $L \times L$ image, we consider the four boxes at the top level of the aforementioned dyadic hierarchy and permute these four boxes randomly. We then apply the same operation independently at the lower levels of the dyadic tree. The probability of shuffling at each level is a free parameter. Crucially, these operations are local in the dyadic tree: they only rearrange the positions of the four children of a parent block, and never mix pixels between different parent blocks at the same level. This guarantees that the multifractal partition sums $Z_q(\ell)$ are invariant due to these shuffles. At the same time, the resulting surrogate image exhibits a visibly different texture, and will generically change the CID (Lempel-Ziv entropy). Fig.~\ref{fig:CID_shuffle} shows the result of such shuffles on the LDOS images obtained from the Hofstadter model at filling $\alpha = 1/8$. We allow the shuffles to be energy dependent. Due to the invariance of $Z_q(\ell)$ under shuffles, the multifractal exponents obtained from the original image (at $E = E_c$) and the corresponding shuffled image are identical, but the CID is clearly quite different, as is evident from Fig.~\ref{fig:CID_shuffle}. The left panels of the figure show the actual original image and the shuffled image at $E = E_c$.

\FloatBarrier
\subsection{Dependence of the LDOS and its CID on the disorder realization}
The simulations in the main text use one configuration of the on-site disorder. Figure \ref{fig:disorder_realizations} repeats them for four independent realizations with all other parameters unchanged ($L=100$ torus, $\alpha=1/10$, $W=0.5\,t$, Gaussian broadening $\eta=0.02\,t$, energy step $0.01\,t$). The CID is computed as in the main text: each LDOS map is averaged to get the mean LDOS and rank-transformed to get CID.

Panel (a) shows $\log\rho(\mathbf r,E)$ at the four subband centres ($E/t = 0.76,\ 1.47,\ 2.36,\ 3.43$) and the three gap midpoints ($E/t = 1.12,\ 1.91,\ 2.89$), one row per realization. The spatial pattern changes from realization to realization, but its character at a given energy does not: fine-grained and space-filling at a subband centre, a few smooth maxima from localized-state tails in a gap. Accordingly, the mean LDOS (b) and the CID of the ranked LDOS (c) are nearly identical for the four realizations. The CID peaks at every subband centre ($0.638$–$0.642$, standard deviation across realizations $3$–$7\times10^{-4}$) and is lowest in the gaps ($0.605$–$0.615$, standard deviation $3$–$5\times10^{-3}$), a contrast of $0.03$–$0.04$, ten to a hundred times the realization-to-realization spread. The energy dependence of the CID is thus set by the Landau-level structure, not by the particular disorder configuration.

\begin{figure}[h!tbp]
     \includegraphics[width=0.95\textwidth]{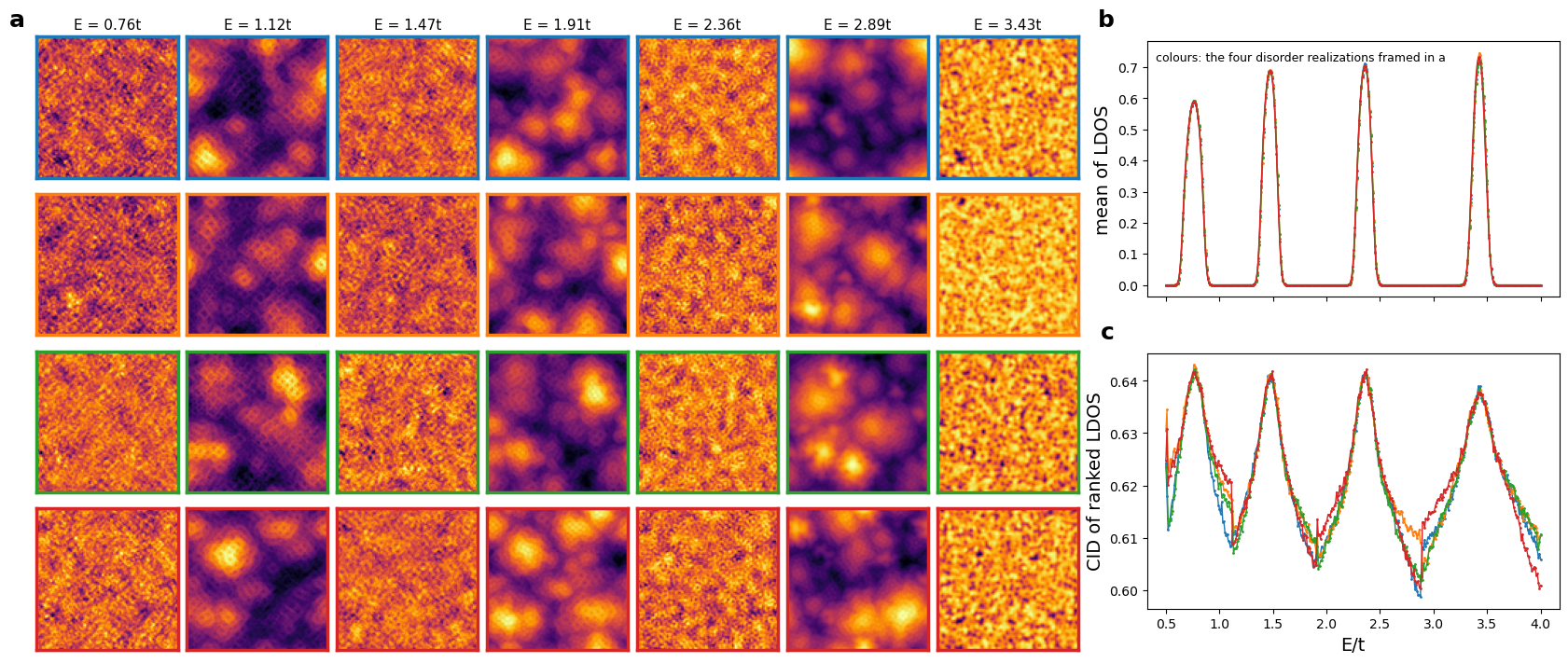}
    \caption{\label{fig:disorder_realizations}
     \textbf{Independence of the LDOS statistics and of the CID from the disorder realization.}
     \textbf{(a)} LDOS maps of the Hofstadter model (log scale, $L=100$, $\alpha=1/10$, $W=0.5\,t$, $\eta=0.02\,t$) for four independent disorder realizations (rows; the frame colour identifies the realization in (b) and (c)) at the four subband centres and the three gap midpoints indicated above the columns. Each map has its own colour scale.
     \textbf{(b)} Mean LDOS versus energy $E$ for the four realizations.
     \textbf{(c)} CID of the rank-transformed LDOS versus $E$ for the same realizations. The CID peaks at every subband centre and is lowest in the gaps for each realization; the spread between realizations ($\lesssim 5\times10^{-3}$) is an order of magnitude smaller than the peak-to-dip contrast.
     }
 \end{figure}

\FloatBarrier
 \subsection{Dependence of  CID on the number of threshold levels}

\begin{figure}[h!tbp]
     \includegraphics[width=0.95\textwidth]{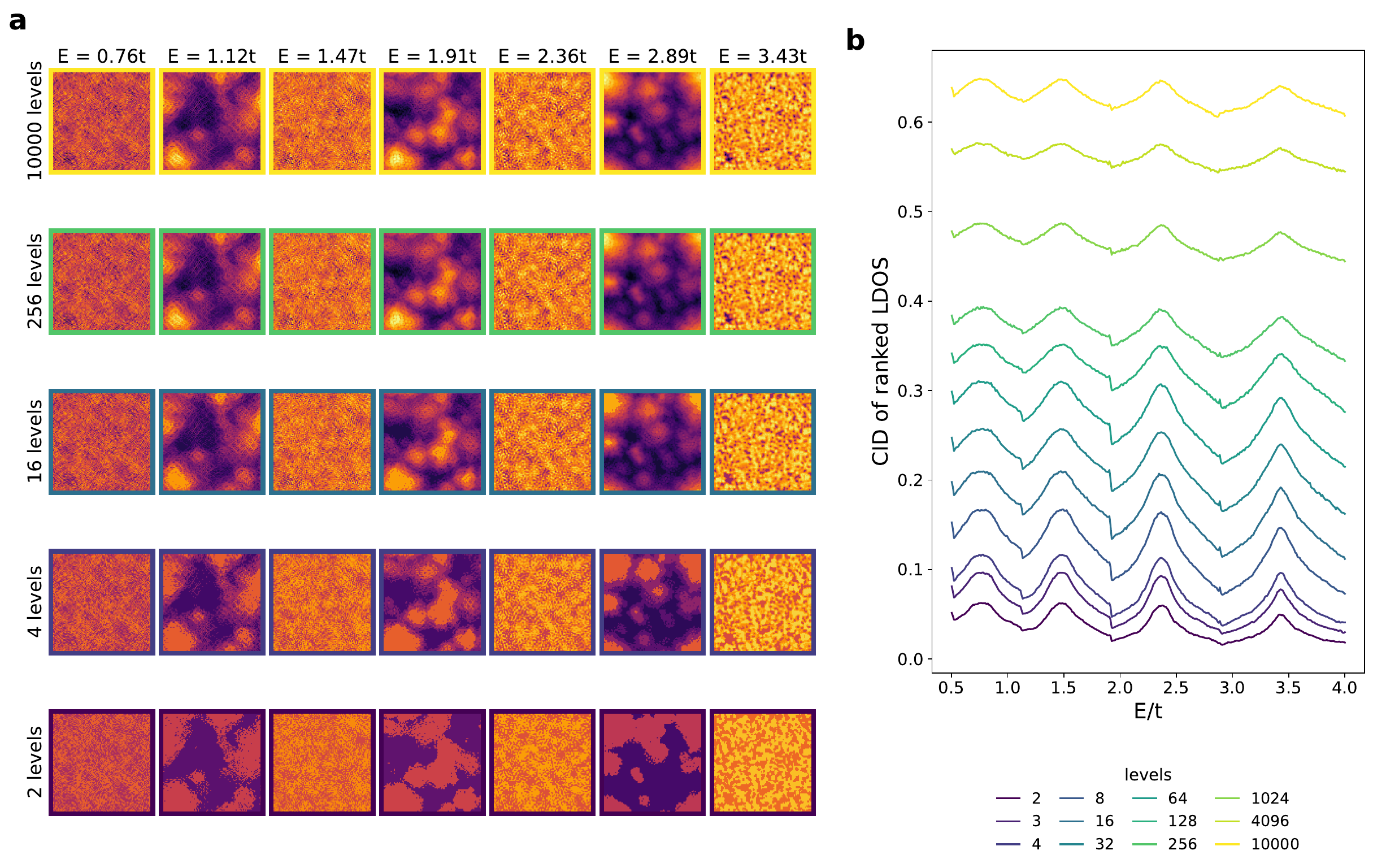}
    \caption{\label{fig:CID_thresholds}
     \textbf{CID of the ranked LDOS as a function of the number of discrete levels. }
Hofstadter model with L = 100, $\alpha$ = 1/10, W = 0.5 $t$ and $\eta$ = 0.02 $t$, for one disorder realization.      \textbf{(a)}
LDOS maps reduced to n threshold levels based on the pixel rank. Columns are subband centres and gap midpoints. Rows go from n = $10^4$ (original map) at the top, down to n = 2, a binary map split at the median. Frame colors match the curves in (b).
   \textbf{(b)} CID versus E/$t$ for n = 2 to $10^4$ levels. Each map is flattened along a Hilbert curve before compression.
     }
 \end{figure}

\textcolor{black}{The CID in Fig.~\ref{fig:disorder_realizations} is computed on the full rank transform of the LDOS, so every pixel keeps its own value.
To check that the energy dependence of the CID does not depend on this choice, we reduce each ranked map to $n$ equally populated levels.
A pixel with rank $r \in \{0,\dots,N-1\}$, where $N = L^2$, is assigned the level $\lfloor r\,n/N \rfloor$.
For $n = N$ (10000 in this case) this gives back the full rank map, and for $n = 2$ it gives a binary map with the threshold set at the median.
Representative maps are shown in Fig.~\ref{fig:CID_thresholds}(a).
At the subband centers the LDOS maps show fine patterns at every $n$, while at the gap midpoints it is made of a few large, smooth domains, which survive the binary reduction.}

\textcolor{black}{Figure~\ref{fig:CID_thresholds}(b) shows the rank-transformed CID as a function of energy for $n$ between 2 and $10^4$.
The overall CID rises with $n$, because more levels carry more information per pixel.
The energy dependence, however, is similar for every $n$.
The relative contrast between peaks and dips is in fact larger for small $n$, since the gap midpoints are easier to compress with smaller $n$.}

\FloatBarrier
\subsection{CID algorithm applied to a site-percolation model: showing the difference from the quantum percolation transition}
To model the emergence of spatial connectivity in a disordered landscape, we implement conventional site percolation and a continuous-threshold growth model on a two-dimensional square lattice of size $1000\times1000$ (the landscape presented in the Fig.~\ref{fig:CID_sitepercolation} is $200\times 200$ for pixel visibility, but $1000\times1000$ is used for the CID and correlation length $\xi$ extraction). In the site-percolation simulation, each lattice site is independently occupied with probability $p$, and connected clusters are identified using nearest-neighbor (four-connectivity) labeling. A percolating state is defined by the presence of a cluster spanning the system from the top to the bottom boundary.

In our simulations the percolation transition occurs at $x \approx 0.23$, corresponding
to an occupation probability $p \approx 0.5933$, in close agreement with the known
threshold $p_c = 0.5927$ for site percolation on the square lattice; near this point
the correlation length $\xi$ shows the expected divergence, cut off by the system size.
The CID, in contrast, is maximal at $p = 1/2$ and shows no feature at $p_c$. This is expected: because the sites are occupied independently, the serialized image is an i.i.d.\ Bernoulli($p$) sequence for any scan order, and the CID converges to its Shannon entropy $H(p) = -p\log_2 p - (1-p)\log_2(1-p)$ \cite{ziv1978compression,wyner1994sliding}, which is maximal at $p = 1/2$ and, by the exact symmetry of the coder under $p \leftrightarrow 1-p$, remains at $p = 1/2$ at every finite size; the curve in Fig.~\ref{fig:CID_sitepercolation}(b) follows $H(p)$ up to a size-dependent offset. The CID of an uncorrelated landscape is therefore a histogram quantity that carries no information about connectivity, even though $\xi$ diverges at $p_c$. In other words, the normalized CID of the site-percolation model will be 1.

To further validate the simulation, we fit the correlation length to the form $\xi \propto |x - x_c|^{-\nu}$ of the parameter $x$, obtaining a critical exponent of approximately $\nu \approx 1.33$, consistent with the theoretical value for two-dimensional percolation. The slight offset between the peaks of CID and the correlation length contrasts with the behavior observed in quantum Hall transitions, reflecting the distinct nature of complexity in classical percolation versus quantum critical systems.

 \begin{figure}[htbp]
     \includegraphics[width=0.9\textwidth]{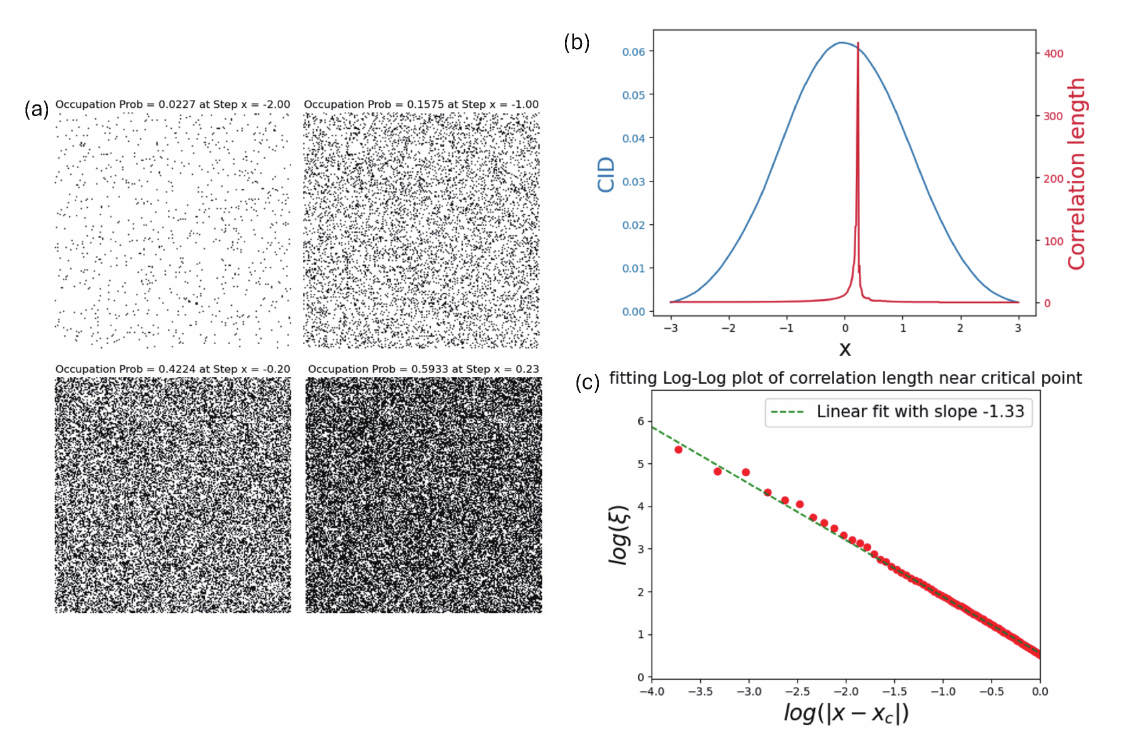}
     \caption{\label{fig:CID_sitepercolation}
    \textbf{CID applied to the site-percolation simulation and extracted correlation length}
(a) Transition all the way to the percolation with occupation probability of 0.5933 .
(b) Calculated CID and the correlation length using second moment estimator.
(c) Critical exponent fit to the correlation length and the driving parameter x.}
 \end{figure}

\subsection{Real space MIM images from another sample region.}
 \begin{figure}[h!tbp]
     \includegraphics[width=0.9\textwidth]{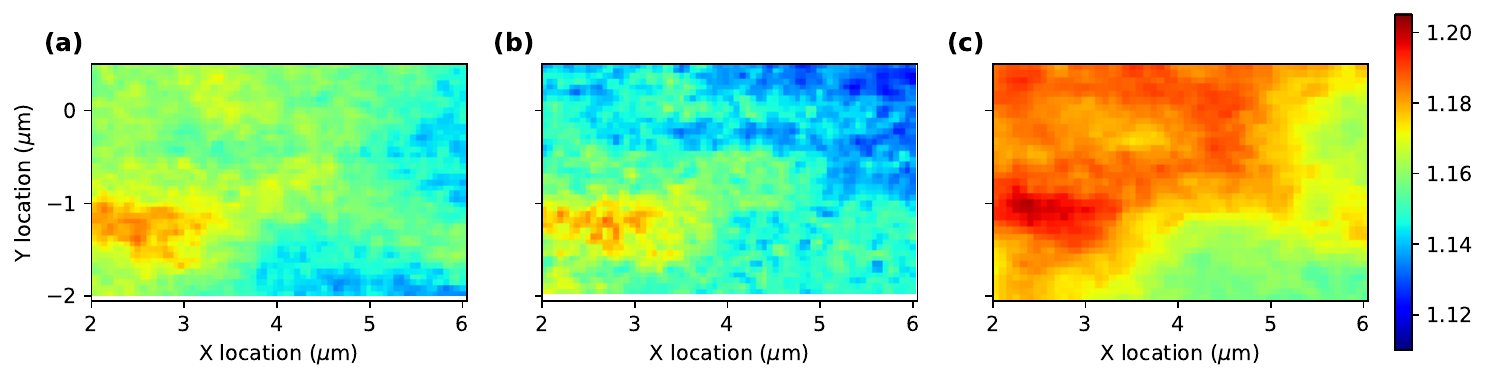}
     \caption{\label{fig:Supp_Stripe}
     \textbf{MIM snapshots in a different sample region, measured at millikelvin temperatures and  B = 8.5 T.  } From left to right:
\textbf{(a)} Transition between the $\nu = 0$ and $\nu = 1/3$ fillings, $V_\mathrm{g} = -1.720$~V.
\textbf{(b)} $\nu = 1/3$ incompressible state, $V_\mathrm{g} = -1.585$~V.
\textbf{(c)} Transition between the $\nu = 2/3$ and $\nu = 1$ fillings, $V_\mathrm{g} = -1.120$~V near the percolation point.
     }
 \end{figure}
In Fig.~\ref{fig:Supp_Stripe} we show representative MIM images acquired in a different sample region at $B = \SI{8.5}{T}$, with the other measurement conditions similar to those used for Fig.~\ref{fig:Simulation}C. The scan area is $4\times2.5~\mu\mathrm{m}^2$. The three panels correspond to gate voltages spanning the transition between $\nu=0$ and $\nu=1/3$ (panel a), the $\nu=1/3$ incompressible state (panel b) and the transition between $\nu=2/3$ and $\nu=1$ (panel c), respectively. Consistent with the MIM snapshots shown in Fig.~\ref{fig:Simulation}C, the transition regimes exhibit spatially extended stripe features associated with compressible regions on top of the incompressible background. These filaments closely resemble those observed in the other sample region, indicating that the spatial patterns reported in the main text are reproducible across different areas of the sample and are not specific to a single scan location.
\FloatBarrier
\subsection{Discussion of the effect of long-range disorder on a quantum Hall system}
\begin{figure}[htbp]
    \includegraphics[width=0.95\textwidth]{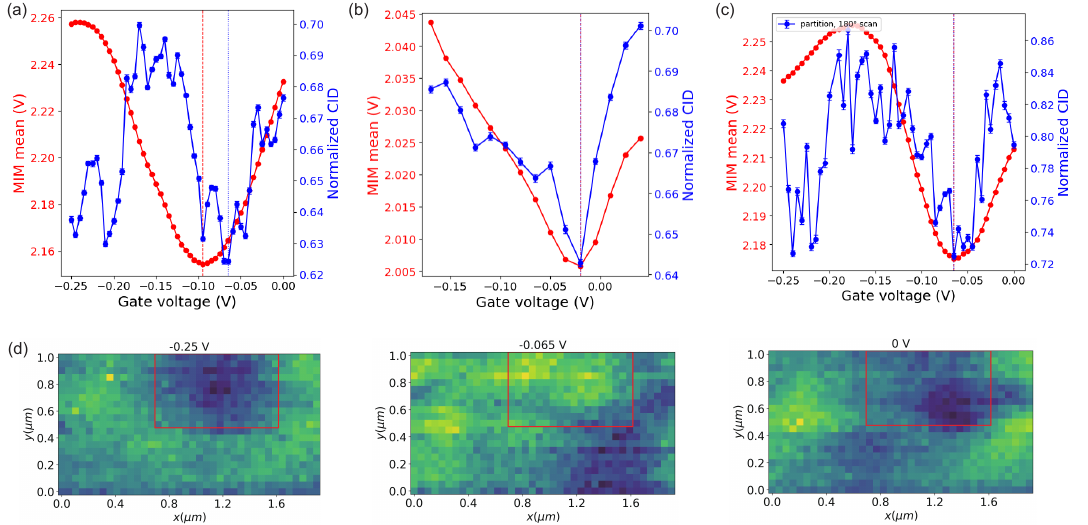}
    \caption{\label{fig:LLOffset}
    \textbf{Effect of long-range disorder on the CID of a double-layer graphene device.}
    \textbf{(a)} Spatially averaged MIM signal (red, left axis) and normalized CID (blue, right axis) of the top graphene layer versus gate voltage, computed over the full scan window. The red and blue dashed lines mark the minima of the averaged MIM signal and of the normalized CID, which do not coincide.
    \textbf{(b)} Same as (a) for the bottom graphene layer, which is screened from surface disorder by the top layer. The two minima approximately coincide (dashed line).
    \textbf{(c)} Same as (a) for the top layer, with the CID computed only within the region marked by the red box in (d), which contains a single dominant disorder feature. The minimum of the normalized CID moves close to that of the averaged MIM signal.
    \textbf{(d)} MIM images of the top layer at $V_\mathrm{g} = -0.25$, $-0.065$ and $0$~V. The red box marks the cropped region used in (c).
}
\end{figure}

We next examine how long-range disorder affects the CID behavior, using the
layer-resolved MIM measurements of a double-layer graphene device reported in
Ref.~\cite{Cao2025layer}. Since the MIM response of the two graphene layers can
be separated, the same Landau-level gap can be compared in a disordered layer and
in a cleaner one. The top graphene layer exhibits a relatively strong and long-range disorder landscape, consistent with its larger carrier-density fluctuations observed in that work. Its spatial texture across the full imaging window is dominated by multiple disorder features, and the normalized CID does not show a clear one-to-one correspondence with the minimum in the averaged MIM response [Fig. \ref{fig:LLOffset}(a)]. In contrast, the bottom graphene layer is substantially cleaner due to screening of the surface disorder by the upper graphene layer, and its normalized CID exhibits a more well-defined minimum that approximately coincides with the minimum of the averaged MIM signal [Fig. \ref{fig:LLOffset}(b)]. To further examine the effect of disorder in the top layer, we restrict the CID analysis to a smaller cropped region containing a single dominant disorder feature, indicated by the red box in Fig. \ref{fig:LLOffset}(d). Within this local region, the normalized CID develops a clearer minimum that roughly coincides with the minimum in the averaged MIM response [Fig. \ref{fig:LLOffset}(c)]. These results indicate that strong, spatially inhomogeneous long-range disorder can obscure the global CID signature, while restricting the analysis to a locally homogeneous region recovers the correspondence between the structural complexity of the MIM landscape and the averaged MIM response.

\FloatBarrier
\section{Applying CID algorithm to other imaging modalities}

\FloatBarrier
\subsection{Discussion of applying CID algorithm to charge density waves in TaS2}

In this subsection we benchmark CID on a time-resolved phase transition: we analyze the ultrafast dark-field transmission electron micrographs of the charge-density-wave (CDW) domain dynamics in a 70~nm 1\textit{T}-TaS$_2$ film reported in Ref.~\cite{danz2021ultrafast}, with data obtained from the associated repository \cite{6GFN9K_2021}. In these images only electrons scattered into the superstructure reflections of the nearly commensurate (NC) CDW reach the detector. NC regions therefore appear bright, while regions in the incommensurate (IC) phase appear dark (Fig.~\ref{fig:TaS2_CDW}(a)). We use the eight frames ($520\times520$ pixels, $\approx4$~nm per pixel) spanning pump-probe delays from $-5$~ps to 3.1~ns. The binary CID using the same NC/IC threshold as defined in the article closely follows the evolution of the domain pattern (Fig.~\ref{fig:TaS2_CDW}(b-c)). It rises from 0.003 before time zero, when the film is homogeneously NC, to a maximum of 0.017 at 0.75--1~ps. It then decreases up to 3.1~ns, when only three small IC islands remain. CID is not, however, a proxy for the IC area itself. At 2~ps and 130~ps the IC fractions are similar (39\% and 44\%), yet CID drops from 0.012 to 0.009. Over this interval the fragmented, speckled IC patches formed shortly after excitation consolidate into a few compact domains with sharp walls, consistent with the domain-wall sharpening reported in Ref.~\cite{danz2021ultrafast}. The binary CID thus reflects the geometric complexity of the phase pattern, set mainly by the total length of the NC/IC domain walls, rather than the fraction of transformed area. The transient increase at 550~ps coincides with the heat-diffusion-driven growth of the IC domains.

\begin{figure}[htbp]
    \includegraphics[width=0.9\textwidth]{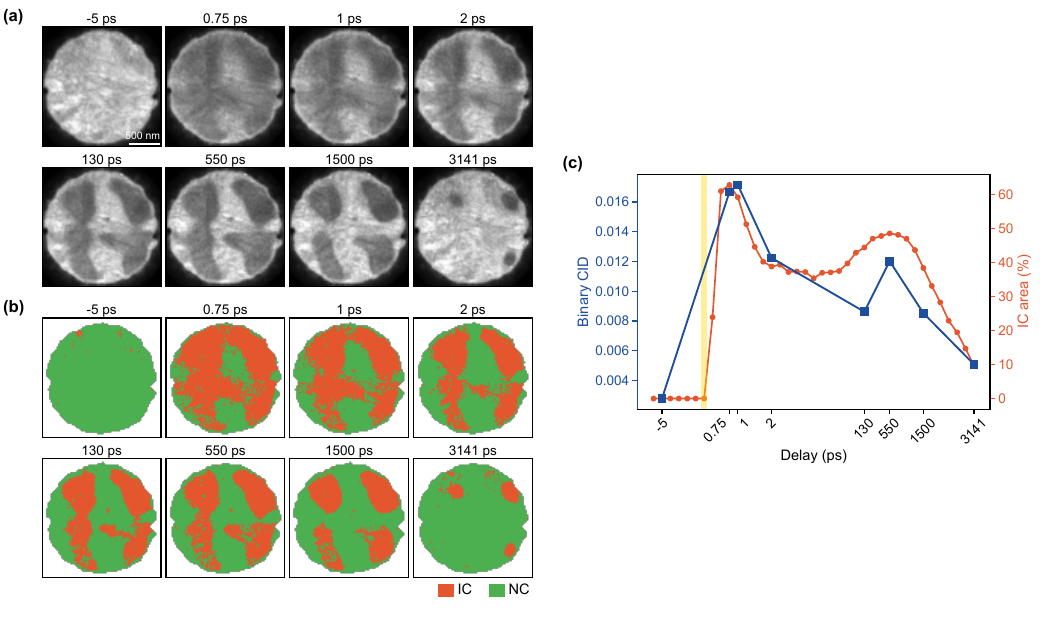}
    \caption{\label{fig:TaS2_CDW}
    \textbf{CID analysis of photo-induced charge-density-wave domain dynamics in 1\textit{T}-TaS$_2$ imaged by ultrafast dark-field electron microscopy.}
    \textbf{(a)} Dark-field micrographs at selected pump-probe delays. Bright and dark contrast correspond to the nearly commensurate (NC) and incommensurate (IC) CDW phases, respectively. Scale bar: 500~nm.
    \textbf{(b)} Binary IC/NC maps obtained by thresholding the images in (a) at a single intensity calibrated to the IC area fraction reported in Ref.~\cite{danz2021ultrafast}. Pixels along the aperture edge are excluded.
    \textbf{(c)} CID of the binary maps (blue, left axis) and the IC area fraction based on the same image series (orange, right axis) versus pump-probe delay. The yellow bar marks the arrival of the pump pulse; the delay axis is not to scale.
    \textcolor{black}{Images in (a) were adapted from the publicly available dataset of Danz \textit{et al.}~\cite{danz2021ultrafast,6GFN9K_2021}, licensed under CC0~1.0; pixels along the aperture edge were excluded, the images were binarized at a single intensity threshold, and they were reanalyzed using the CID procedure described here.}
    }
\end{figure}

Next we apply the same binary CID procedure to the steady-state experiment of Ref.~\cite{danz2021ultrafast}, in which the film is heated by a continuous-wave laser whose intensity is swept up and then back down (Fig.~\ref{fig:TaS2_CW}(a)). A single IC domain nucleates near the film center, grows until it nearly fills the aperture, and shrinks again on the down-sweep (Fig.~\ref{fig:TaS2_CW}(b)). With the threshold calibrated to the reported transformed (IC) area, the binary CID tracks the IC area across all frames (Fig.~\ref{fig:TaS2_CW}(c)). It also reproduces the hysteresis of the first-order transition: at the same heating intensity, the down-sweep lies above the up-sweep. Because only one compact domain is present, its wall length and its area grow together. CID and IC area therefore coincide here, unlike in the pump-probe series, where the domain geometry and the transformed fraction evolve independently.

\begin{figure}[h!tbp]
    \includegraphics[width=0.9\textwidth]{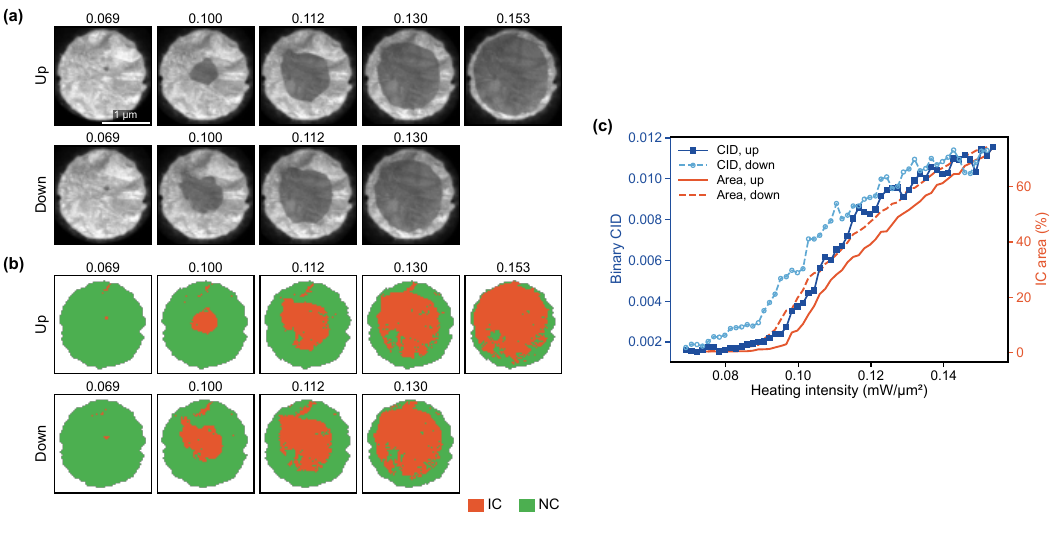}
    \caption{\label{fig:TaS2_CW}
    \textbf{CID analysis of the NC--IC phase transition in 1\textit{T}-TaS$_2$ under continuous-wave laser heating.}
    \textbf{(a)} Dark-field micrographs at selected heating intensities (in mW/$\mu$m$^2$) during the up-sweep (top row) and down-sweep (bottom row). Scale bar: 1~$\mu$m.
    \textbf{(b)} Binary IC/NC maps obtained by thresholding the images in (a) at a single intensity.
    \textbf{(c)} CID of the binary maps (blue, left axis) and the IC area (orange, right axis) versus heating intensity for the up-sweep (solid lines) and down-sweep (dashed lines).
    \textcolor{black}{Images in (a) were adapted from the publicly available dataset of Danz \textit{et al.}~\cite{danz2021ultrafast,6GFN9K_2021}, licensed under CC0~1.0; the images were binarized at a single intensity threshold and reanalyzed using the CID procedure described here.}
    }
\end{figure}
\FloatBarrier
\subsection{Applying CID algorithm to XMCD-PEEM imaging: magnetic domain evolution across the AF–FM transition in FeRh}

FeRh undergoes a first-order metamagnetic transition from an antiferromagnetic (AF) to a ferromagnetic (FM) state slightly above room temperature. In XMCD-PEEM the AF phase gives no dichroic contrast, while FM domains appear as regions of bright or dark contrast. We applied CID to XMCD-PEEM images of a 5 µm FeRh island recorded on heating from 317 to 411 K (Fig.~\ref{fig:AFFM}(a)) and on cooling from 413 to 304 K (Fig.~\ref{fig:AFFM}(b)), taken from Temple \textit{et al.} \cite{temple2018antiferromagnetic}, with data obtained from the associated repository \cite{Temple2018FeRhDataset}. Each image was quantized to 64 uniform levels over a symmetric range common to the whole series. Raw CID (Fig.~\ref{fig:AFFM}(c)) rises across the transition, but this mainly reflects the growing dichroic signal spreading the pixel values over more quantization levels rather than a change in domain structure. Once this histogram contribution is removed, the normalized CID (Fig.~\ref{fig:AFFM}(d)) shows the opposite trend, decreasing from about 0.75 in the AF state to 0.63 at 411 K and recovering on cooling. This decrease arises from the FM phase organizing into large, uniformly magnetized domains. CID therefore captures the evolution of domain geometry across the transition without segmentation or a calibrated order parameter, and could serve as a complementary indicator in imaging modalities where no such quantity is readily available.

\begin{figure}[htbp]
    \includegraphics[width=0.8\textwidth]{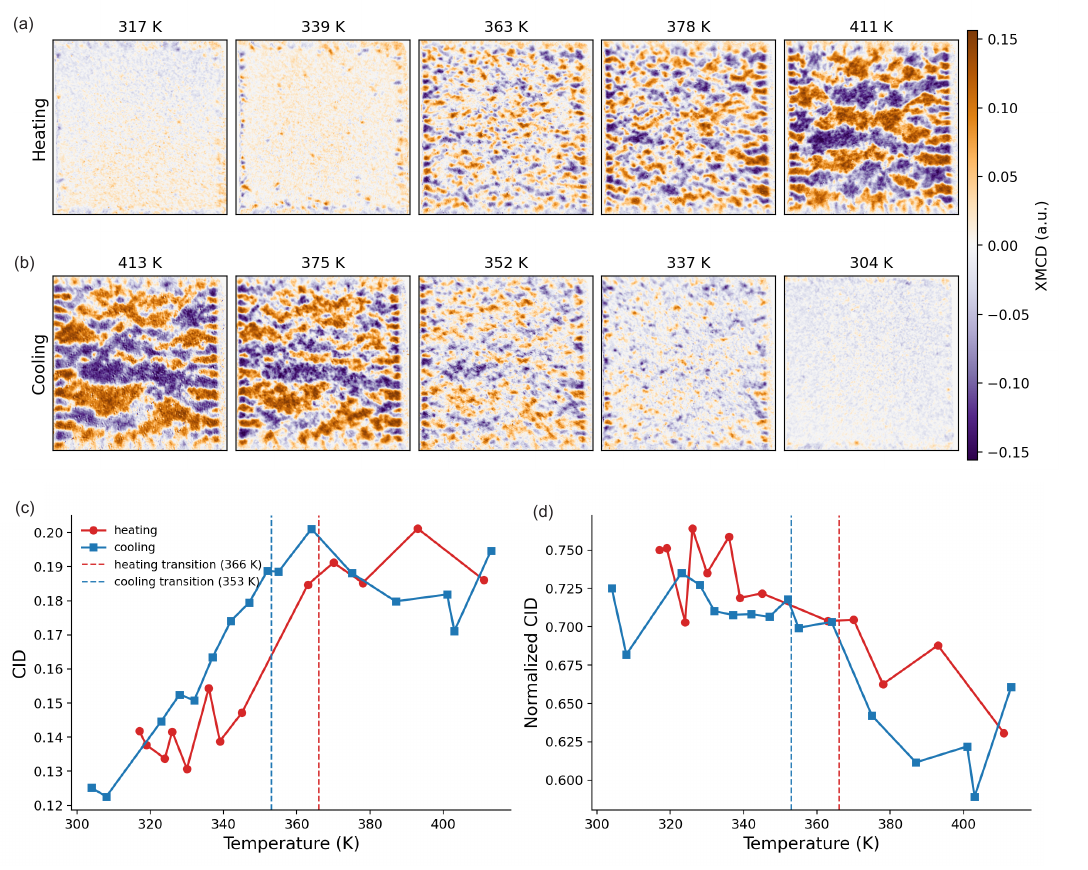}
    \caption{\label{fig:AFFM}
    \textbf{CID analysis of XMCD-PEEM imaging of FeRh: tracking magnetic domain evolution across the antiferromagnetic--ferromagnetic transition.}
    \textbf{(a, b)} XMCD-PEEM images of a 5~$\mu$m FeRh island on (a) heating and (b) cooling, on a common color scale. The AF phase appears white; orange and purple FM domains are magnetized parallel and antiparallel to the X-ray beam.
    \textbf{(c)} CID and \textbf{(d)} normalized CID versus temperature on heating (red circles) and cooling (blue squares). Dashed lines mark the heating (366~K) and cooling (353~K) transition temperatures, determined from the XMCD signal as defined in Ref.~\cite{temple2018antiferromagnetic}. \textcolor{black}{Images in (a) were adapted from the publicly available dataset of Temple \textit{et al.}~\cite{temple2018antiferromagnetic,Temple2018FeRhDataset}, licensed under CC~BY~4.0; the images were cropped to the island, quantized to 64 uniform levels, displayed on a diverging color scale, and reanalyzed using the CID procedure described here.}
    }
\end{figure}
\FloatBarrier
\subsection{Applying CID algorithm to STXM imaging of magnetic skyrmions: tracking field- and current-driven changes in skyrmion textures}

In this subsection we apply the CID algorithm to the raw STXM images of current-driven skyrmions in a Pt/CoB/Ir multilayer wire published by Zeissler \textit{et al.}~\cite{zeissler2020diameter} (Fig.~\ref{fig:CID_skyrmion}), with data obtained from the associated repository \cite{Zeissler2019SkyrmionDataset}. The normalized CID of the images (0.70--0.79) responds to both control parameters of the experiment: it shows an overall increase from $\approx 0.71$ at 0~mT to $\approx 0.75$ at $-4$~mT, despite frame-to-frame fluctuations (Fig.~\ref{fig:CID_skyrmion}(c)), consistent with the reported shrinking and annihilation of skyrmions, and it drops with nearly the same initial decrease for positive and negative current pulses (Fig.~\ref{fig:CID_skyrmion}(d)), consistent with the rapid reorganization and partial coalescence of the initially nucleated skyrmion texture. The measure also applies to non-image data: for the count maps of the skyrmion Hall angle versus diameter and velocity from the same work (Fig.~\ref{fig:CID_skyrmion}(b)), the normalized CID of 0.76--0.92 captures the clustering of the events around $\theta_{\mathrm{Sky}} \approx 10^\circ$ at every field, with the rise towards $-4$~mT reflecting mainly the smaller number of events there (165 at 0~mT, 40 at $-4$~mT). CID thus captures field- and pulse-driven changes in the underlying data structure without modality-specific tuning.

\begin{figure}[h!tbp]
    \centering
    \includegraphics[width=0.85\textwidth]{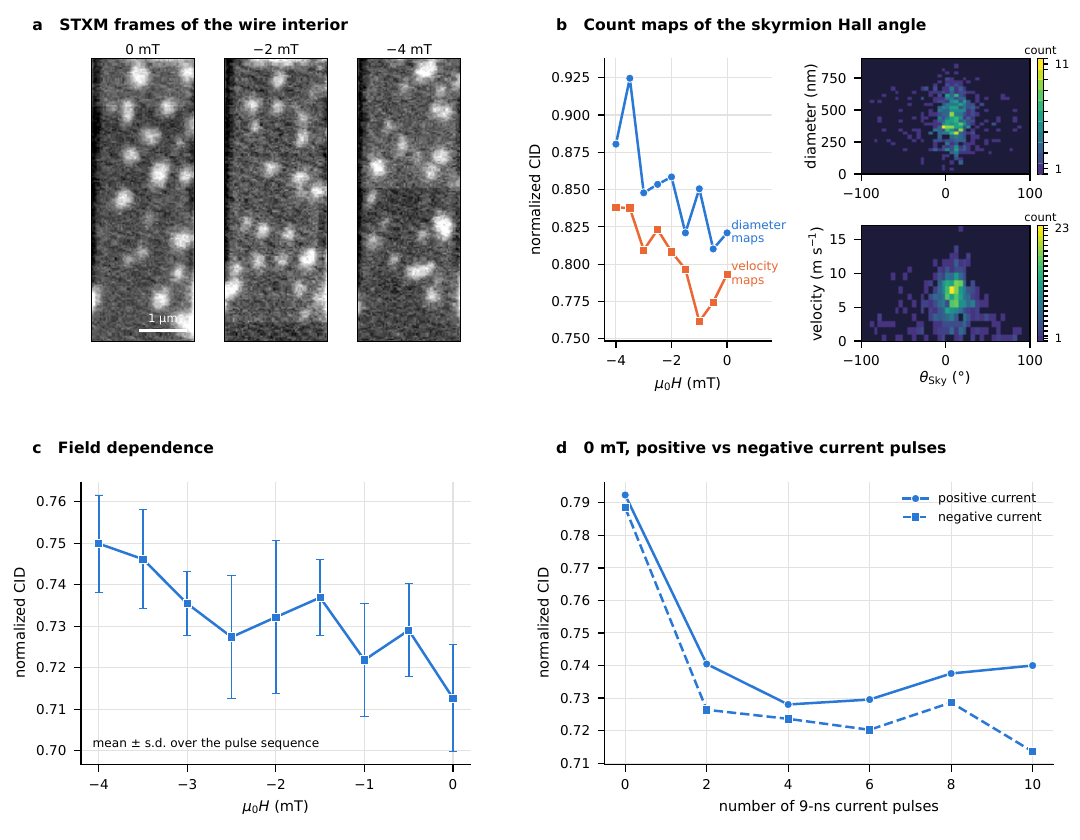}
    \caption{\label{fig:CID_skyrmion}
    \textbf{Normalized CID of scanning transmission X-ray microscopy (STXM) images of magnetic skyrmions in a chiral multilayer wire.}
    \textbf{(a)} Example frames at 0, $-2$ and $-4$~mT out-of-plane field (bright: skyrmions).
    \textbf{(b)} Count maps of the skyrmion Hall angle $\theta_{\mathrm{Sky}}$ versus skyrmion diameter (25~nm $\times$ $5^\circ$ bins) and versus velocity (1~m/s $\times$ $5^\circ$ bins) for all fields (680 displacement events; the per-field maps of Ref.~\cite{zeissler2020diameter}, Figs.~3 and 4), and the normalized CID of the per-field maps (no smoothing; 165 events at 0~mT decreasing to 40 at $-4$~mT).
    \textbf{(c)} Field dependence of the normalized CID, mean $\pm$ s.d.\ over the 8--11 frames of each sequence.
    \textbf{(d)} Zero-field series after successive positive and negative current pulses (frames of Ref.~\cite{zeissler2020diameter}, Fig.~1d--o). Scale bar in (a): 1~$\mu$m.
    \textcolor{black}{Images in (a) were adapted from the publicly available dataset of Zeissler \textit{et al.}~\cite{zeissler2020diameter,Zeissler2019SkyrmionDataset}, licensed under CC~BY~4.0; the STXM frames were reanalyzed using the CID procedure described here, and the count maps in (b) were recomputed from the published skyrmion-event data.}
    }
\end{figure}
\FloatBarrier
\subsection{Applying CID algorithm to optical imaging of biofilms: tracking wrinkle pattern coarsening, arrest and refinement}

To test CID on a biological system, we analyze phase-contrast time-lapse images of wrinkling Pseudomonas aeruginosa biofilms from Geisel \textit{et al.} \cite{geisel2022role}, with data obtained from the associated repository \cite{Geisel2021BiofilmDataset}. All frames cropped to the channel interior and quantized to 64 levels by rank. The first dataset (Fig.~\ref{fig:CID_biofilm}(a)) follows the three stages of wrinkle formation defined in Ref.~\cite{geisel2022role}: isolated wrinkles appear in an initially flat film (stage 1), merge into a connected network (stage 2), and the network then stabilizes (stage 3). The normalized CID (Fig.~\ref{fig:CID_biofilm}(b)) decreases from 0.93 to 0.80 through stages 1 and 2 and levels off in stage 3. The flat film initially contains only fine, uncorrelated texture, which is nearly incompressible. As wrinkles appear and widen in stage 1, the pattern coarsens, and in stage 2 they join into long, continuous ridges that compress well. In stage 3 the network layout is essentially fixed and the feature size barely changes, so CID stays constant.
To test whether CID also captures a reversal of this trend, we analyze a second time-lapse (Fig.~\ref{fig:CID_biofilm}(c)) in which the nutrient medium was replaced by a salt solution at 55 h and restored at 73.5 h. During the salt treatment, growth is arrested and the pattern stays frozen, so the normalized CID (Fig.~\ref{fig:CID_biofilm}(d)) remains near 0.84. Once nutrients are restored, new thin wrinkles form across the film and the characteristic feature size shrinks. This finer, less ordered pattern is harder to compress, and the normalized CID rises to about 0.88 within 6 h. As these wrinkles merge and the network coarsens again, the normalized CID decreases to about 0.85. In both experiments, CID therefore tracks the evolution of the wrinkle pattern: it stays constant when the pattern is fixed, falls as it coarsens, and rises when finer structure appears.

\begin{figure}[h!tbp]
    \includegraphics[width=0.8\textwidth]{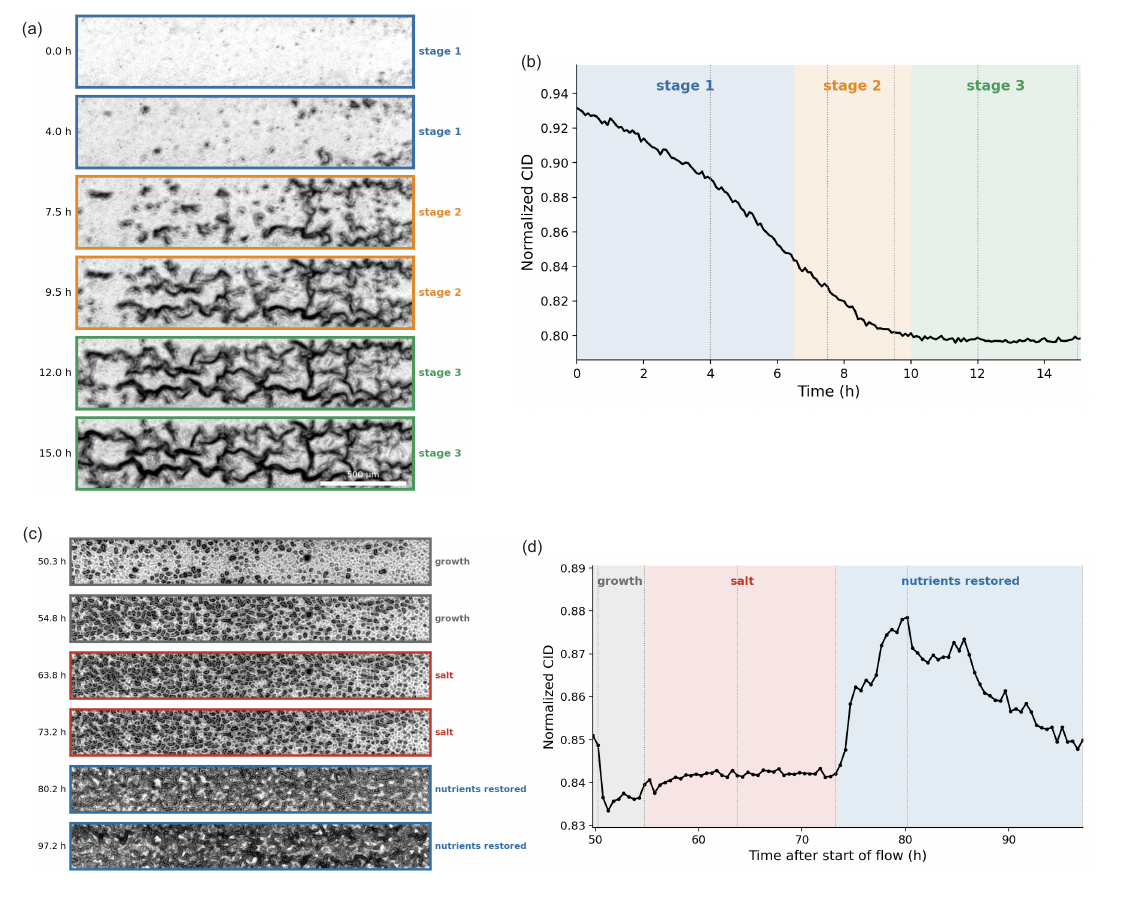}
    \caption{\label{fig:CID_biofilm}
    \textbf{CID analysis of biofilm wrinkling: tracking pattern coarsening, arrest and refinement. }
    \textbf{(a)} Phase-contrast images of the biofilm at 3 stages as defined in Ref.~\cite{geisel2022role}.
    \textbf{(b)} Normalized CID versus time; shaded regions mark the three stages and dotted lines mark the times shown in (a).
    \textbf{(c)} Cropped phase-contrast images of the biofilm at representative times during growth (gray), salt treatment (red) and after nutrients were restored (blue).
    \textbf{(d)} Normalized CID versus time; shaded regions mark the three phases and dotted lines mark the times shown in (c).
    \textcolor{black}{Images in (a) and (c) were adapted from the publicly available dataset of Geisel \textit{et al.}~\cite{geisel2022role,Geisel2021BiofilmDataset} (Dryad), licenced under CC0~1.0; the images were cropped to the channel interior, quantized to 64 levels by rank, and reanalyzed using the CID procedure described here.}
    }
\end{figure}

\FloatBarrier
\section{Coarse-graining and image-size dependence of CID}
\FloatBarrier
\subsection{Coarse-graining of the simulated LDOS}
To mimic the finite spatial resolution of an SPM experiment, we check the CID at different coarse-grained image sizes to capture the resolution limitation in the SPM tip (Fig.~\ref{fig:CID_coarse_grained}). The simulated image contains $100\times 100$ pixels  which, according to the magnetic length, has the real space size of 800nm. The MIM tip size is around 75nm which defines the resolution. So an effective $10\times 10$ pixel representation is comparable to the spatial resolution of the experiment. We coarse-grained the simulated space into different sizes from 90 pixels down to 10 pixels using the Gaussian weighted renormalization and observe that the overall CID profile is preserved through this down-sampling process.
\begin{figure}[htbp]
    \includegraphics[width=0.7\textwidth]{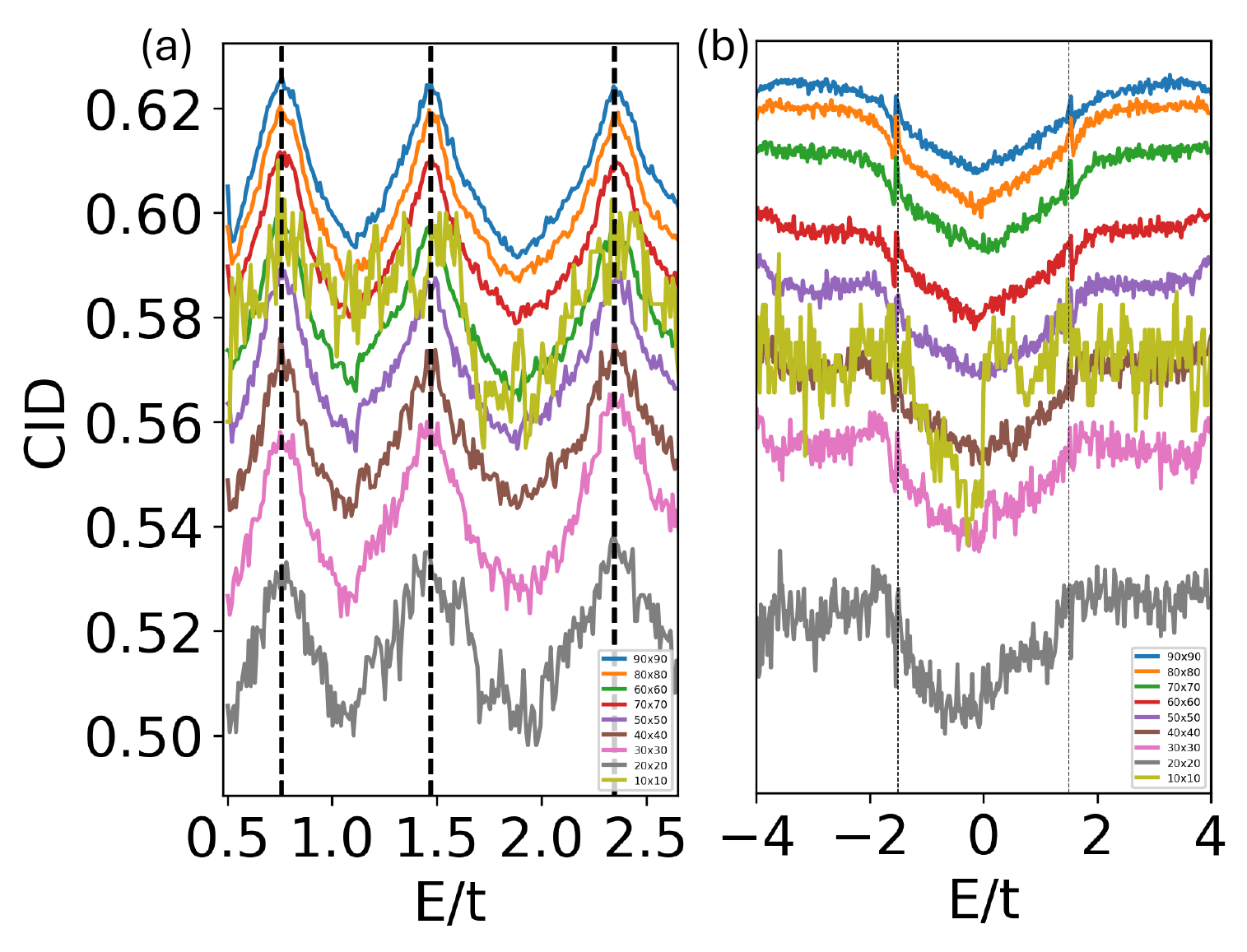}
    \caption{\label{fig:CID_coarse_grained}
    \textbf{CID of the coarse-grained LDOS.}
    The CID calculation of the (a) simulated QHE and (b) simulated band insulating state based on the coarse-graining pixel numbers from 90 to 10 pixels using Gaussian weighted renormalization. }
\end{figure}

The $10\times 10$ pixels will be a little bit small for the CID algorithm because the LZ77 compressor will tend to add headers or other incompressible components which now become comparable with the data information. Even so, we could still see the peak aligned with the results from larger pixels.

\subsection{Scale dependence of CID: a benchmark on SEM images}
The CID of an image depends on how the image parameters (the pixel size, spatial resolution of the probe, and the number of pixels) as well as on the
pattern itself. The main text compares CID values only within series of fixed
size, due to the limitation of the homogeneous sample area. Here we characterize the dependence on a simple test sample: Scanning Electron Microscope (SEM) images of a lithographically patterned
periodic structure. The
analysis also verifies the applicability claim in the main text by applying the same
CID procedure to another imaging modality.

We use three SEM images of the same structure taken at nominal
magnifications of $166\times$, $504\times$ and $1.10\text{k}\times$ (Fig.~\ref{fig:CID_coarse_grained_SEM}(a), first row), each $8192\times6144$ pixels and step sizes $\delta = 84.2$, $27.7$ and $12.7$~nm. CID is computed
exactly as for the MIM data: each image is serialized along a Hilbert curve (Sec.~\ref{sec:serialization}) and losslessly compressed with LZ77. We reduce the pixel
numbers by block-averaging: $b\times b$ blocks are averaged to one pixel
and rounded to the nearest integer, coarsening the effective pixel to $b \times\delta$ while
keeping the field of view fixed. The down-sampled representative images at $b=1$ (raw), 16,
64 and 256 are shown in Fig.~\ref{fig:CID_coarse_grained_SEM}(a), corresponding to effective pixel sizes of $0.01$--$0.08~\mu$m,
$0.20$--$1.35~\mu$m, $0.81$--$5.39~\mu$m and $3.25$--$21.5~\mu$m (top to bottom;
ranges span the three magnifications). The calculated CID values as a function of the block size collapse onto one non-monotonic curve, agreeing to
within $\approx0.015$ over $0.15\lesssim b\times\delta\lesssim2~\mu$m. CID rises from native
resolution to a maximum of $0.29$--$0.31$ at $b\times\delta\approx0.9$--$1.2~\mu$m, then
falls.

We interpret the rise (orange band, $0.1$--$1.1~\mu$m) as the removal of
redundancy between over-sampled adjacent pixels, and the fall (blue band,
$1.1$--$8~\mu$m) as decreasing structure: The  channels and bars visible
in the top rows of Fig.~\ref{fig:CID_coarse_grained_SEM}(a) are being smoothed away once the averaging window exceeds
the length scale. At large $b$ the down-sampled images themselves become small,
and CID rises again for two reasons: 1) There is no more regular information that can be compressed, as the image becomes more random; 2) From the LZ77 algorithm, a short compressed sequence
carries a fixed per-stream overhead (header/checksum) that is no longer negligible
relative to the payload. We mark points with downsampled image width $\leq32$~px
as finite-size (open symbols, dotted lines in Fig.~\ref{fig:CID_coarse_grained_SEM}(b) and exclude
them from the physical interpretation above; the rise past this point is a property of
the compression, not of the sample.

\begin{figure}[htbp]
    \includegraphics[width=0.8\textwidth]{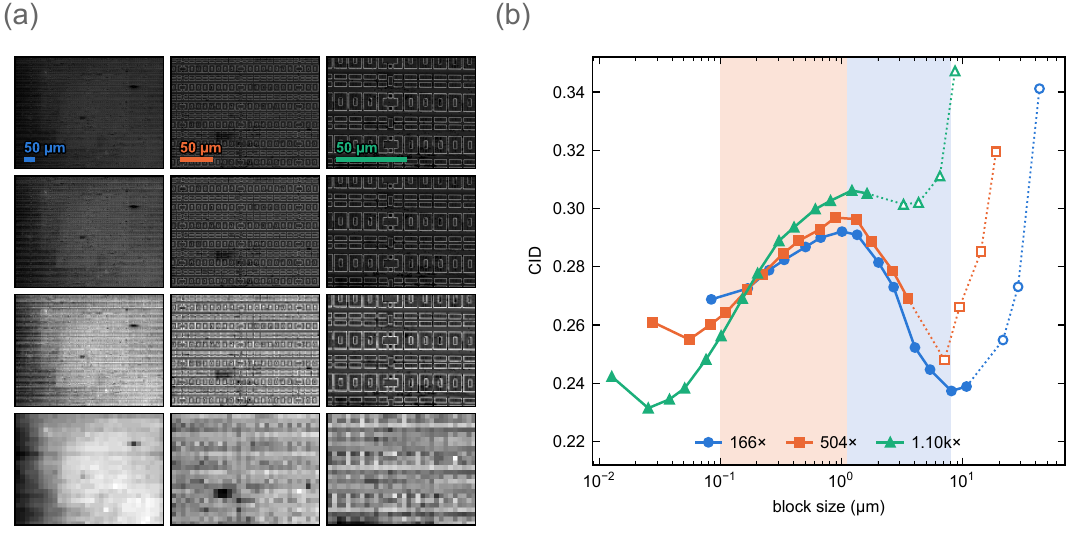}
    \caption{\label{fig:CID_coarse_grained_SEM}
    \textbf{CID of the coarse-grained SEM image. }
      (a) Block-averaged images at $b=1$, 16, 64 and 256 (rows, top to bottom) for
  the three magnifications $166\times$, $504\times$ and $1.10\text{k}\times$ (columns;
  scale bars 50~$\mu$m). The effective pixel size after averaging is $b\times\delta$, ranging
  from $0.01$--$0.08~\mu$m (top row) to $3.25$--$21.5~\mu$m (bottom row) across the
  three magnifications. The grayscale display images are auto-scaled for better visualization.
  (b) CID versus physical block size $b\delta$ for the same three images (colors as in
  (a)). Maximum CID appears between the orange band ($0.1$--$1.1~\mu$m, where averaging removes
  oversampling redundancy) and the light-blue band ($1.1$--$8~\mu$m, where averaging
  begins to smooth away real spatial structure). Open symbols and dotted lines mark
  points with image width $\leq32$~px, where the finite length of the
  compressed sequence dominates. }
\end{figure}

Absolute CID values depend on the physical
pixel size relative to the finest resolved structure in the image, so comparisons are
meaningful only within series acquired at fixed sampling -- which is how every CID
series in the main text was acquired.
The MIM frames in the main text contain approximately 30–40 pixels per dimension and therefore lie near the crossover where finite-size compression effects become appreciable.
Within a series of equal-size images this algorithmic offset is approximately common to all images, so the
relative comparisons in the main text remain interpretable, even though absolute CID values should not
be compared across image sizes.

\FloatBarrier
\subsection{Coarse-graining study on MIM images of an antiferromagnetic semiconductor}

\begin{figure}[htbp]
    \includegraphics[width=0.85\textwidth]{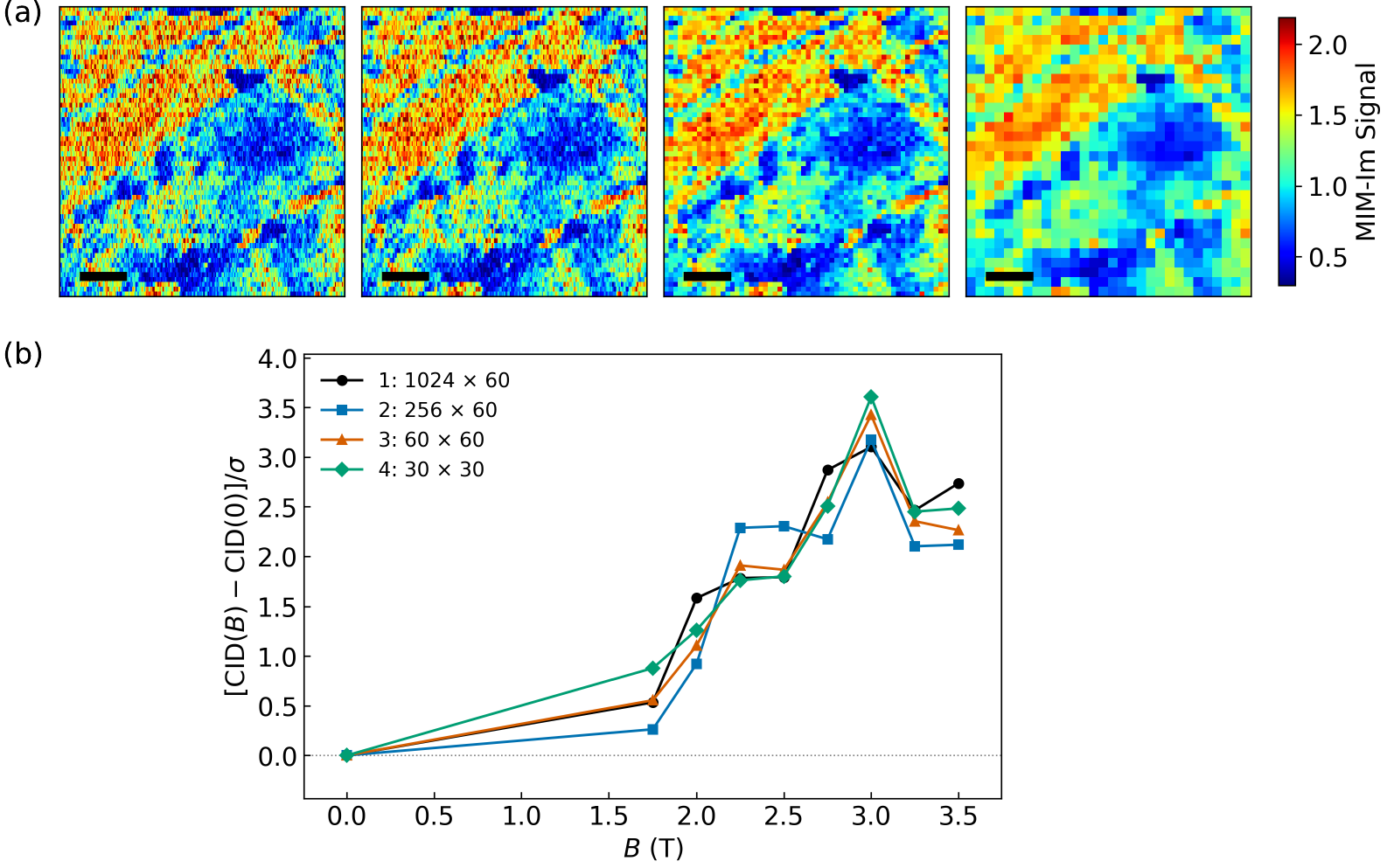}
    \caption{\label{fig:CID_Eu526}
    \textbf{CID of the coarse-grained MIM spatial snapshot of Eu$_5$Sn$_2$As$_6$. }(a) MIM-Im image of Eu$_5$Sn$_2$As$_6$ (data reported in \cite{day2025colossal};
    MIM excitation frequency 3~GHz, $T=100$~mK; Scan range $30\times30~\mu$m$^2$, with $1024\times60$ pixels). From left to
    right: block sizes along $x$ of $0.03$ (raw), $0.12$, $0.5$ and $1~\mu$m.
    Scale bars: 5~$\mu$m.
    (b) Scaled CID, $[\mathrm{CID}(B)-\mathrm{CID}(0)]/\sigma$, versus magnetic field for
    the four coarse-graining levels show in (a), where $\sigma$ is the standard deviation of each
    curve over $B$. The overall field dependence, a steep rise between $1.75$ and
    $2.25$~T and a maximum at $3$~T, is preserved at all block sizes.
    }
\end{figure}
The SEM benchmark above shows that CID first rises as block averaging
removes redundancy between over-sampled pixels. For SEM the probe dimension (electron beam) is
far smaller than any pixel considered, so that redundancy only comes from the sample geometry.
In a scanning-probe measurement, adjacent pixels smaller than the tip size also
give largely the same signal, as the tip apex averages over a finite area. Here we
repeat the coarse-graining analysis on an MIM image that is over-sampled along one
axis, and examine how the field dependence of CID changes with the number of pixels.
We select the MIM-Im image of the antiferromagnetic semiconductor Eu$_5$Sn$_2$As$_6$ acquired at a magnetic field of $3.5$ T \cite{day2025colossal}. The image covers
$30\times30~\mu$m$^2$ with $1024$ pixels along the fast-scan axis $x$
($\delta_x = 30$~nm) and $60$ lines along $y$ ($\delta_y = 0.5~\mu$m), which shows pronounced spatial contrast.

Representative MIM-Im images at different coarse-graining factors are shown
in Fig.~\ref{fig:CID_Eu526}(a), which contain $1024\times60$ (raw), $256\times60$, $60\times60$, and $30\times30$ pixels,
respectively, from left to right.
Fig.~\ref{fig:CID_Eu526}(b) shows scaled CID versus magnetic field $B$ for the four coarse-graining levels in Fig.~\ref{fig:CID_Eu526}(a). Because the absolute CID depends on the pixel number and sampling (demonstrated in Fig.~\ref{fig:CID_coarse_grained_SEM}), each raw CID curve is subtracted from its zero-field value and scaled by the standard deviation over $B$, i.e., $[\mathrm{CID}(B)-\mathrm{CID}(0)]/\sigma$. The detailed field dependence slightly varies with block size, but the overall trend is preserved at all block sizes down to $30\times 30$ pixels: CID increases steeply up to $2.25$~T and reaches a maximum around $3$~T. The field-driven evolution of spatial complexity therefore remains identifiable under moderate coarse-graining, although finer features of the curve depend on the sampling.

In summary,  the coarse-graining analysis addresses two aspects of sampling: the pixel size and the number of pixels. For the SEM
images (Fig.~\ref{fig:CID_coarse_grained_SEM}), the initial rise continued up to $\sim1~\mu$m, the scale of the patterned structure.
Below the probe resolution, finer pixels cannot resolve new information, so the finest meaningful pixel size for CID in a scanning-probe image is set by the larger of the probe and the sample feature resolution. The MIM frames in the main text use a $75-100$~nm pixel width, comparable to the
tip resolution: they are not
over-sampled to speed up measurements.
As for the number of pixels, although the magnitude of the CID variation depends strongly on sampling, its overall field dependence persists down to $30\times30$ pixels.
 Together this section demonstrates that frames of the size around 30 pixels are sufficient in capturing the essential spatial features and their evolution with the tuning parameter.

\FloatBarrier
\section{Applying CID to device fabrication imaging}
\FloatBarrier
\subsection{Applying CID algorithm to sequence of optical images}

In this subsection we apply it to a sequence of optical images recorded during the pickup of hBN flakes, in which a stamp was used repeatedly to remove residual flake material from the substrate (Fig. \ref{fig:CID_pickup}). Across the seven scans in this sequence, the flake-covered area fraction decreases steadily from 8.5$\%$ to 5.1$\%$ as material is progressively removed. The normalized CID, computed from the illumination-corrected images, decreases correspondingly from 0.47 to 0.44 over the sequence. This behavior confirms that the CID metric captures the reduction in spatial heterogeneity of the flake surface as the pickup process proceeds, illustrating the applicability of the method to optical image analysis more broadly.

 \begin{figure}[htbp]
     \includegraphics[width=0.95\textwidth]{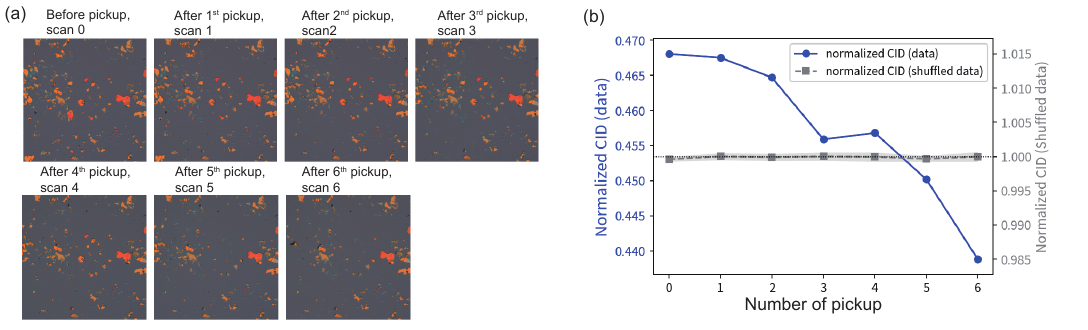}
     \caption{\label{fig:CID_pickup}
    \textbf{CID analysis of hBN pickup optical images}
(a) Optical images of hBN flakes during a stamp-based pickup process, ordered by scan number (0–6).
(b) Normalized CID computed from images in (a), plotted as a function of scan number.}
 \end{figure}

\FloatBarrier
\subsection{Applying CID algorithm to AFM cleaning of surface residue}

Here we apply CID to a series of atomic force microscopy (AFM) lateral force images acquired during mechanical cleaning of residual surface contamination via repeated contact-mode AFM scanning (Fig. \ref{fig:CID_afmclean}). Across six sequential scans, residue features visible in the lateral signal are progressively swept away, leaving a markedly cleaner surface, and the normalized CID decreases correspondingly. This behavior is consistent with the sensitivity of the CID method to surface structure observed for the other datasets in this work, and further supports its applicability across imaging modalities.

 \begin{figure}[htbp]
     \includegraphics[width=0.95\textwidth]{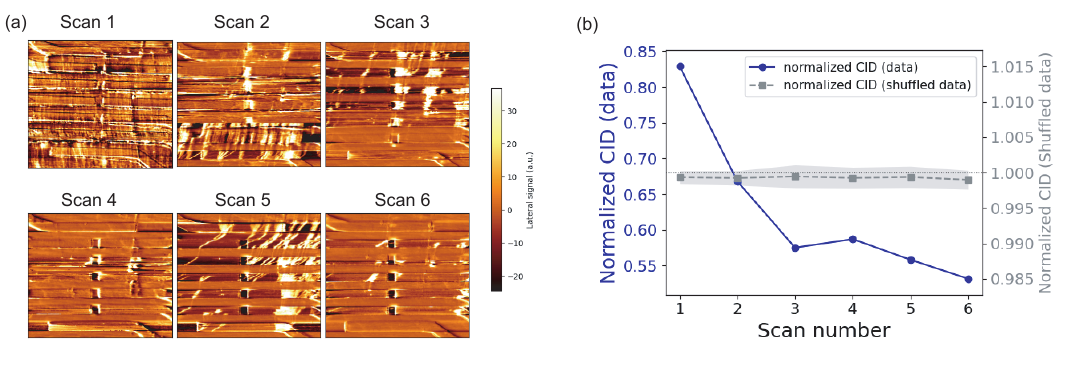}
     \caption{\label{fig:CID_afmclean}
    \textbf{CID analysis of an AFM-based surface cleaning process}
(a) Lateral force AFM images (Scan 1–6) acquired during repeated contact-mode scanning used to mechanically remove residual surface contamination. (b) Normalized CID (blue, left axis) computed from the images in (a), and the corresponding shuffled-data control (gray, right axis), plotted as a function of scan number. The normalized CID decreases across the six scans as residue is removed.
}
 \end{figure}

\stopcontents

\end{widetext}

\end{document}